\def\eg{{\it e.g.}}%
\def\etal{{\it et~al.}}%
\def\Journal#1#2#3#4{{#1} {\bf #2}, #3 (#4)}

\def\NIM{\em Nucl. Instrum. Methods}

\def\NPB{{\em Nucl. Phys.} B}
\def\NPO{{\em Nucl. Phys.}}
\def\PLB{{\em Phys. Lett.}  B}
\def\PLO{{\em Phys. Lett.}}
\def\PRL{\em Phys. Rev. Lett.}
\def\PRD{{\em Phys. Rev.} D}
\def\PR{{\em Phys. Rev.}}
\def\ZPC{{\em Z. Phys.} C}
\def\EPJ{{\em Eur. Phys. Jour.} C}
\def\CPC{{\em Comput. Phys. Comm.}}
\def\HS#1{\hspace*{#1truecm}}
\def\VS#1{\vspace*{#1truecm}}
\newcommand{\phoo}{\phantom{00}}

\def\rts {\ensuremath{\sqrt{s}}}

\def\g1z{$g^Z_1$}

\def\HFF{g_{\rm H\FF}}
\def\HEE{g_{\rm H\EE}}
\def\HBB{g_{\rm H\BB}}
\def\HZSM{g_{\rm ZZH}^{\rm SM}}
\def\HbSM{g_{\rm \BB}^{\rm SM}}

\def\hZZ{g_{\rm ZZh}}

\def\hBB{g_{\rm h\BB}}
\def\ABB{g_{\rm A\BB}}
\def\hAZ{g_{\rm ZhA}}
\def\ra{\ensuremath{\rightarrow}}

\def\lapprox{\ensuremath{\sim\kern-1em\raise 0.65ex\hbox{$<$}}}
\def\rapprox{\ensuremath{\sim\kern-1em\raise 0.65ex\hbox{$>$}}}
\def\GeV{\ifmmode {\mathrm{\ Ge\kern -0.1em V}}\else
                   \textrm{Ge\kern -0.1em V}\fi}%
\def\MeV{\ifmmode {\mathrm{\ Me\kern -0.1em V}}\else
                   \textrm{Me\kern -0.1em V}\fi}%
\def\keV{\ifmmode {\mathrm{\ ke\kern -0.1em V}}\else
                   \textrm{ke\kern -0.1em V}\fi}%
\def\eV{\ifmmode  {\mathrm{\ e\kern -0.1em V}}\else
                   \textrm{e\kern -0.1em V}\fi}%

\def\MZ{{\rm m_Z}}
\def\MW{\rm m_W}
\def\MH{\rm m_H}
\def\Mh{\rm m_h}
\def\MA{\rm m_A}
\def\MTOP{\rm m_t}
\def\GZ{\Gamma_{\rm Z}}
\def\gf{\ensuremath{G_{\mathrm{F}}}}

\def\EE{\rm e^+e^-}
\def\MM{\mu^+\mu^-}
\def\TT{\tau^+\tau^-}
\def\LL{\rm \ell^+\ell^-}
\def\FF{\rm f\,\bar{f}}
\def\QQ{\rm q\overline{q}}

\def\BB{\rm b\overline{b}}

\def\ZZ{\rm ZZ}
\def\WW{\rm W^+W^-}

\def\TOP{\rm t\overline{t}}

\def\FFG{\FF(\gamma)\ }
\def\HZ{\rm Z\,H} 
\def\HF{\rm H\FF}
\def\EEZH{\rm e^+e^-\rightarrow Z\,H}
\def\EEhZ{\rm e^+e^-\rightarrow h\,Z}
\def\EEhA{\rm e^+e^-\rightarrow h\,A}

\def\zhllbb{\rm ZH\rightarrow \LL\BB}
\def\zhqqbb{\rm ZH\rightarrow \QQ\BB}
\def\zhnnbb{\rm ZH\rightarrow \nu\bar{\nu}\BB}
\def\zhqqtt{\rm ZH\rightarrow \QQ \TT}

\def\zhllww{\rm ZH\rightarrow \LL\WW}
\def\zhqqww{\rm ZH\rightarrow \QQ\WW}
\def\HTOX{\rm H\rightarrow X}
\def\NUNUH{\rm \nu\bar{\nu} H}

\def\elljet{\LL+2$-$\rm jet}

\def\HTOF{\rm H\rightarrow \FF}
\def\HTOB{\rm H\rightarrow \BB}
\def\HTOV{\rm H\rightarrow VV}
\def\HTOW{\rm H\rightarrow \WW}

\def\HTOG{\rm H\rightarrow \gamma \gamma}
\def\HTOGL{\rm H\rightarrow gg}

\def\elljet{\LL+2$-$\rm jet}

\def\ZMM{\rm Z\rightarrow\MM}

\def\EEEE{\EE\rightarrow\EE}
\def\EEMM{\EE\rightarrow\MM}
\def\EETT{\EE\rightarrow\TT}
\def\EEFF{\EE\rightarrow \FF\ }
\def\EEQQ{\EE\rightarrow\QQ}

\def\EELL{\EE\rightarrow\LL}
\def\EETOP{\EE\rightarrow\TOP}

\def\EFFG{\EE\rightarrow\FFG}

\def\EEWW{\EE\rightarrow\WW}
\def\EEZZ{\EE\rightarrow\ZZ}

\def\EEWEN{\EE\rightarrow {\rm W^+ e^-\overline{\nu}}}

\documentclass[12pt,twocolumn]{article}
\usepackage{epsfig}

\begin{document}
\begin{titlepage}
\Large
\begin{flushright}
DESY 01-218 \\
hep-ex/0202007
\end{flushright}
\vspace{2.8cm}

\LARGE
\begin{center}
{$\EE$ Physics at LEP and a Future Linear Collider
\normalsize\footnote{
Based on lectures given at the International School-Seminar
'Actual Problems 
of Particle Physics' in Gomel, Belarus.}}

\vspace{.8cm}
\Large
Wolfgang Lohmann \\

{DESY Zeuthen}
\end{center}
\Large\centerline{December 7, 2001}

\vspace{1.8cm}
\LARGE
\abstract{
A summary of results obtained from $\EE$ 
annihilations at LEP is given.
The precision measurements around the Z resonance,
the results from charged gauge boson production
and 
searches for new particles are reviewed.
Particular emphasis is devoted to the Higgs boson.
The prospects of an $\EE$ linear collider in the energy range
of about 1 TeV are discussed.
}

\end{titlepage}

\stepcounter{section}
\section*{\large \arabic{section}.\ \large Introduction}

Annihilations of electrons and positrons
were a long time ago in particular the testing ground
of Quantum Electrodynamics (QED),~\eg~to test the
'pointlikeness' of leptons in the couplings to the photon
\cite{pointl}. 
The first hints that electron interactions 
can proceed not only via a virtual photon
came from neutrino scattering, 
having shown evidence for neutral currents~\cite{nc}. 
These were interpreted
as the exchange of a heavy neutral gauge boson Z.
The discovery of the Z, together with the
discovery of charged gauge bosons
W$^{\pm}$~\cite{wzdisc} was a great step to give credibility
to the Standard Model (SM)~\cite{sm}, unifying
the electromagnetic and weak interactions.  
The LEP accelerator~\cite{lep} was designed to
test the SM with high precision.
The accuracy finally reached allowed the confirmation
of the SM on the level of quantum corrections
in $\EE$ annihilations on the Z resonance.
After upgrading of the LEP accelerator
with superconducting cavities, the
beam energies were increased above the W-pair production threshold, and
self couplings of gauge 
bosons~\cite{hagiwa} were studied. 
This new energy domain was accessible for the first time, and the data
are
used to search for the missing key-stone of the
SM, the Higgs boson~\cite{higgs}, and for unexpected phenomena. 
Since neither the Higgs boson nor new particles were found
limits on their masses are derived.

The precision measurements at LEP favour a Higgs boson
with a mass below 200 \GeV. The ideal machine
to study such a Higgs boson and to explore 
the mechanism of spontaneous symmetry breaking
is a linear $\EE$ collider in the energy range
up to one TeV. The physics potential of the TESLA
project~\cite{tesla} is discussed. 

\stepcounter{section}
\section*{\large \arabic{section}.\ \large LEP Accelerator and Experiments}

The LEP accelerator, the largest facility
for particle physics research, was taken into operation
in 1989. The accelerator and storage ring had an circumference 
of about 27 km. Four experiments, called ALEPH, DELPHI, L3 and OPAL
~\cite{lepex}
were placed equidistant along the ring. In the first stage, from 1989
to 1995, electron and positron beams were accelerated
to an energy of about 45 \GeV~and the Z resonance was measured 
with high precision. The peak luminosity of the machine reached about
10$^{32}\rm{cm^{-2}s^{-1}}$.
Each of the experiments recorded about 5$\times 10^6$ Z decays.
 After 1995 the energy was enlarged in several steps 
to more then 100 \GeV~ per beam, exploring a new energy domain
of $\EE$ annihilations.       

\stepcounter{section}
\section*{\large \arabic{section}.\ \large The Basic Process}

\begin{figure}
\epsfig{file=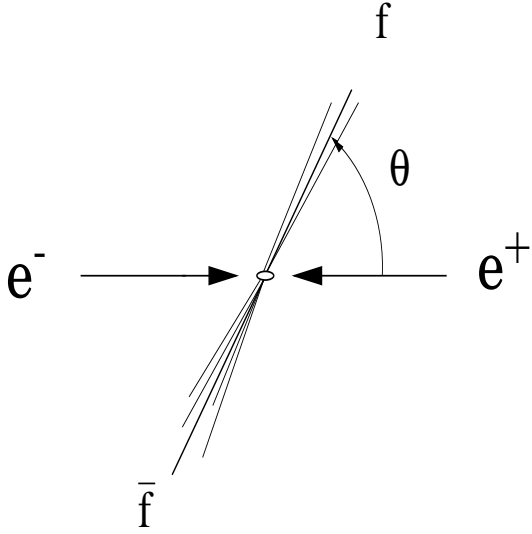,width=7cm, height=7cm}
\caption{Schematic view of the $\EE$ annihilation
into a fermion-antifermion pair. In case the fermions are
quarks they hadronise into jets.}
\label{fig:epem1}
\end{figure}

\begin{figure}[htb]
\begin{center}
    \includegraphics[width=0.4\textwidth,height=0.2\textheight]{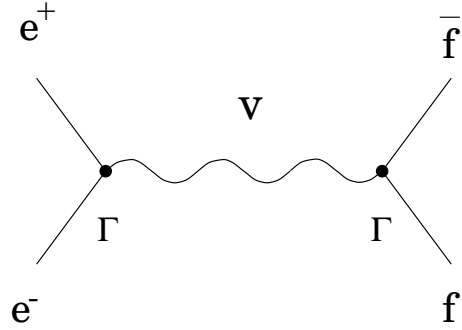}
    \caption[]{\label{fig:epemfg}
              The Feyman-graph for  $\EE$ annihilation.
              V is the exchanged vector boson. The vertex
              functions are denoted as $\Gamma$.    
            }
\end{center}
\end{figure}

The basic process is the 
annihilation of electrons and positrons
into a fermion-antifermion pair, as 
depicted in Figure~\ref{fig:epem1}.
In the case of a symmetric accelerator
the laboratory system is the centre-of-mass
system and  the total energy of the annihilation, $E_{tot}$, is:
\begin{eqnarray*} 
{E_{tot}} = 2{E_b} = \sqrt s, 
\end{eqnarray*} 
where $E_b$ is the beam energy and $s$ the square of the sum
of the four momenta
of the electron and the positron, $p_{e^-}$ and $p_{e^+}$,
\begin{eqnarray*} 
s = (p_{e^-}+p_{e^+})^2
  = 4 E_b^2.
\end{eqnarray*}
The polar angle of the outgoing fermion, 
$\theta$, is  obtained from the momenta of the outgoing fermion and the 
beam electron, $\wp_f$ and $\wp_{e^-}$,
\begin{eqnarray*}
cos \theta = \frac{1}{|\wp_f|\cdot|\wp_{e^-}|}(\wp_f\cdot\wp_{e^-}).
\end{eqnarray*}
The lowest order Feynman-graph for $\EE$ annihilations
into $\FF$
is shown in  
Figure~\ref{fig:epemfg}. 
In QED the exchanged gauge boson V is
the photon and the vertex function $\Gamma$ equals to $Qe\gamma^\mu$,
where $Q$ is the charge of the fermion, $e$ the unit charge and
$\gamma^{\mu} $
are the Dirac matrices.
The matrix element:
\begin{eqnarray*} 
M \propto \frac{Qe^2}{s}(\bar{u}^e \gamma_\mu u^e)(\bar{u}^f \gamma^\mu u^f),
\end{eqnarray*}
where $u^f$ are the four component spinors,
 leads then to the differential and total cross sections
\footnote{For annihilations into leptons $Q$=1.}
\begin{eqnarray*}
\frac{d\sigma}{d\Omega} = \frac{\alpha^2}{4s}(1 + cos^2\theta)
\end{eqnarray*}
and
\begin{eqnarray*}
\sigma = \frac{4 \pi \alpha^2}{3 s}.
\end{eqnarray*}
Here $\alpha$ is the 'fine structure' or electromagnetic
coupling constant, $\alpha = e^2 /4\pi$.

Assuming, instead of the photon, the exchange of a heavy neutral 
boson Z,
the propagator term is replaced by a Breit-Wigner
function. Allowing in addition
a more general Lorentz structure, the
vertex function is extended to 
 $\Gamma = \gamma^\mu(v_f-a_f\gamma^5)$, introducing vector,
$v_f$, and axial-vector, $a_f$, couplings 
of the neutral current.
The matrix element reads then
\begin{eqnarray*} 
M &\propto&  \frac{Qe^2}{s}\chi(s)(\bar{u}^e\gamma_\mu(v_e-a_e\gamma^5)u^e) \\
  &       &  (\bar{u}^f\gamma^\mu(v_f-a_f\gamma^5)u^f),
\end{eqnarray*}
where $\chi(s)$ is the Breit-Wigner function
\begin{eqnarray*}
\chi(s) = \frac{-\gf \MZ^2 s}{2\sqrt{2}\pi\alpha(s-\MZ^2+i\Gamma_Z \MZ)}.
\end{eqnarray*}
Here $\gf$ is the Fermi constant,
 $\MZ$ the mass and $\Gamma_Z$ the width of
the Z boson.
The cross sections obtained, allowing both photon and 
Z exchange, are:
\begin{eqnarray*}
\frac{d\sigma}{d\Omega} & =& \frac{\alpha^2}{4s}(R(s)(1 + \cos^2\theta)+ \\
                        &  &  I(s)\cos \theta)
\end{eqnarray*}
and
\begin{eqnarray}
\sigma = \frac{4 \pi \alpha^2}{3 s} R(s)
\end{eqnarray}
with, in the case of leptons \footnote{for quarks this has to be multiplied
by a colour factor $N_c=3$ and the quark charge must taken into account
in the photon couplings.},

\begin{eqnarray*} 
R(s)& =& 1 + 2v_e v_f \Re (\chi(s)) +   \\
    &  &  (v_e^2+a_e^2)(v_f^2+a_f^2)|\chi(s)|^2   \\
I(s)& =&4 a_e a_f \Re (\chi(s)) +  \\
    &  &8 v_e v_f a_e a_f |\chi(s)|^2. 
\end{eqnarray*}
The term $I(s)$ results from the interference of photon and Z and vector and
axial-vector contributions to the Z exchange.

\stepcounter{section}
\section*{\large \arabic{section}.\ \large Standard Model parameters}

The Lagrangian of the SM obeys 
the $SU(2)\otimes U_Y(1)$
gauge symmetry, giving rise to two gauge couplings, $g'$ and $g$,
as free parameters. Two additional free parameters result from the 
spontaneous symmetry breaking, the
vacuum expectation value
of the Higgs field, $\upsilon_0$, and the parameter
$\lambda$~determining the Higgs potential. 
These quantities 
are related to  
couplings usually measured in the experiment:
\begin{eqnarray*}
\alpha &=& \frac{1}{4 \pi} \frac{g^2g'^2}{g^2+g'^2} = \frac{1}{137.0359895} \\
\gf    &=& (\upsilon_0^2\sqrt 2)^{-1} \\
       &=& 1.166389 \times 10^{-5} \GeV^{-2}        \\
\MZ &=& \frac{\sqrt{g^2+g'^2}}{2} \upsilon_0  .
\end{eqnarray*}  
The measurement of the three quantities $\alpha$, $\gf$ and $\MZ$
would allow us to determine three free parameters of the theory.
The couplings $\alpha$ and $\gf$ are measured
in other experiments with errors of
$4.5\times 10^{-8}$~\cite{codata} 
and $1.9\times 10^{-5}$~\cite{pdg}, respectively.
The measurement of $\MZ$ is the missing link at this point, and performed 
by the LEP experiments.

\stepcounter{section}
\section*{\large \arabic{section}.\ \large Mass and Width of the Z}

\begin{figure}[htb]
  \begin{center}
    \includegraphics[width=0.53\textwidth]{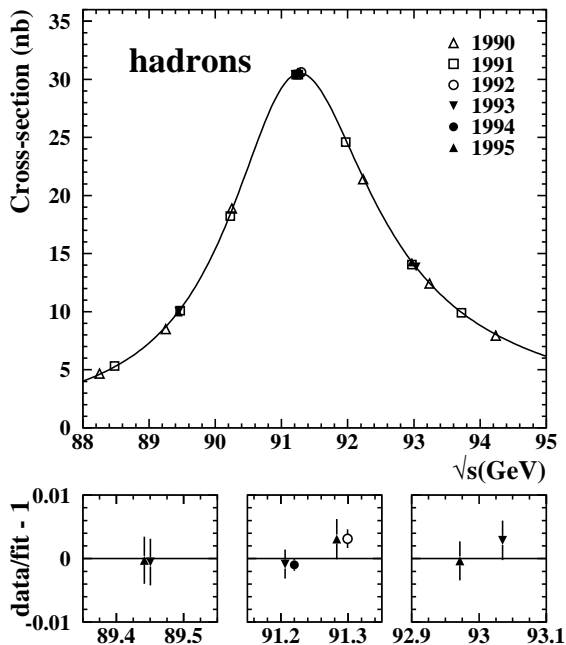}
    \caption[]{\label{fig:stot}
              The cross section for  $\EE$ annihilations
              into hadrons as function of the energy
              as measured by the OPAL collaboration.
              Also shown in the bottom plot is the deviation
              between data and fit. 
            }
  \end{center}
\end{figure}

\begin{figure}[htb]
  \begin{center}
    \includegraphics[width=0.5\textwidth]{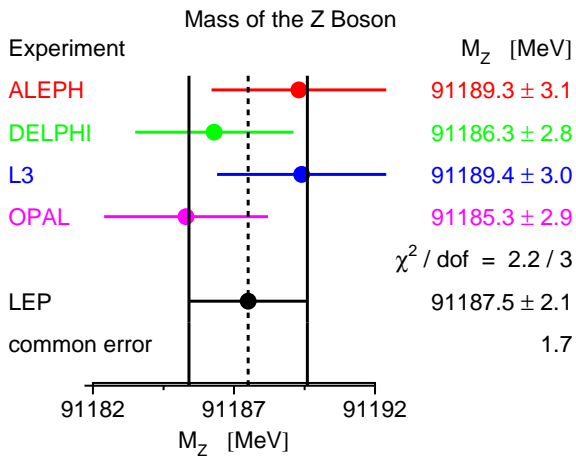}
    \caption[]{\label{fig:mz}
              The results of the Z mass measurement by the
              four LEP experiments and their combination.
            }
  \end{center}
\end{figure}

In order to measure the mass and width of the Z, $\MZ$ and $\GZ$,
we look first on the cross section 
for Z exchange only. It can be written
as:
\begin{eqnarray*}
\sigma_{\rm Z} = 12\pi\frac{\Gamma_{ee}\Gamma_{f\bar{f}}}{\MZ^2}
              \frac{s}{(s-\MZ^2)^2+(\frac{s\Gamma_Z}{\MZ^2})^2}.
\end{eqnarray*}
The partial width, $\Gamma_{f\bar{f}}$, is related to the vector and
axial-vector couplings:
\begin{eqnarray*}
\Gamma_{f\bar{f}}=\frac{G_F \MZ^3}{6\pi \sqrt2}(a_f^2+v_f^2).
\end{eqnarray*}
The measurement of $\MZ$ can be
performed by measuring the cross
section as function of $\sqrt s$ and performing a fit using 
Eqn.(1), where $\sigma_{\rm Z}$ is the
dominant contribution in the resonance region.
Bremsstrahlung, mainly
off the initial state electrons, and other higher order corrections,
which change the shape of the cross section, have to be
taken into account. The theoretical framework
used to describe the energy dependence of the cross section is given in
the software codes 
ZFITTER~\cite{zfitter} and TOPAZ0~\cite{topaz}.\\
As an example, the result from the OPAL
experiment~\cite{opalz} for hadronic final states 
is given in Figure~\ref{fig:stot}.
The measurements of the cross section as function
of $\rts$ is well described
by the fit using Eqn.(1),
including radiative
corrections,   
with $\MZ$, $\GZ$ and the partial widths as free
parameters.
These measurements were done
for the processes $\EELL$, with $\ell = {\rm e}$, $\mu$ or $\tau$,
and $\EE \ra$ hadrons  
by all LEP experiments. 
The results for $\MZ$ are shown
in Figure~\ref{fig:mz}~\cite{lepew}.
They are in good agreement 
and combined equal to $\MZ = 91187.5\pm2.1$ \MeV.
In the combination errors common to all experiments, \eg
~due to the uncertainty of the LEP beam energy,
are taken into account. 
The measurement
of $\MZ$ has a precision of $2.2\cdot 10^{-5}$, 
well comparable with
the one of the Fermi constant $\gf$. 
The measurements of $\GZ$ are also combined
from all LEP experiments yielding 
$\GZ$ = 2.4952$\pm$0.0023 \GeV.
Using the partial widths, $\Gamma_{f\bar{f}}$,
the width of invisible
decays, $\Gamma_{inv}$, is obtained as 
$\Gamma_{inv}$ = 499.0 $\pm$ 1.5 \MeV.
Within the SM this quantity is interpreted in terms 
of the number of neutrino species. The result is
$N_\nu =$ 2.984 $\pm$ 0.008.
 
\stepcounter{section}
\section*{\large \arabic{section}.\ \large Couplings of the Z}

In the SM the 
vector and axial-vector
couplings of the Z to fermions are predicted as:
\begin{eqnarray*}
v_f &=&  I_3-2 Q_{\rm f} s_W^2 \\
a_f &=&  I_3.
\end{eqnarray*}
Here $I_3$ is the third component of the weak isospin,
$Q_{\rm f}$ the charge of the fermion
and $s_W$ the sinus of the electroweak mixing angle
$\Theta_W$.
A prove of the SM is to compare these predictions to
measurements.

The measurements 
are obtained using the total cross section, where
$\sigma_Z$ depends on $a_f$ and $v_f$, the
differential cross section $d\sigma/d\cos\theta$
and polarisation asymmetries.

Defining the 
forward-backward asymmetry, ${\cal A}_{FB}$, as the normalised
difference of the cross sections integrated between $-1< \cos \theta
<0$ and $0 < \cos \theta < 1 $, respectively, one obtaines
on the peak of the resonance:
\begin{eqnarray*}
{\cal A}_{FB}&=&  \frac{\sigma_F-\sigma_B}{\sigma_F+\sigma_B}   \\
             &=&\frac{3}{4}\frac{v_e a_e}{v_e^2+a_e^2} \frac{v_f a_f}{v_f^2+a_f^2}  \\
             &=&\frac{3}{4} {\cal A}_e {\cal A}_f .
\end{eqnarray*}
From the measurement of ${\cal A}_{FB}$ the product ${\cal A}_e {\cal A}_f$
can be determined.

\begin{figure}[htb]
\hspace{-0.5cm}
    \includegraphics[width=0.58\textwidth]{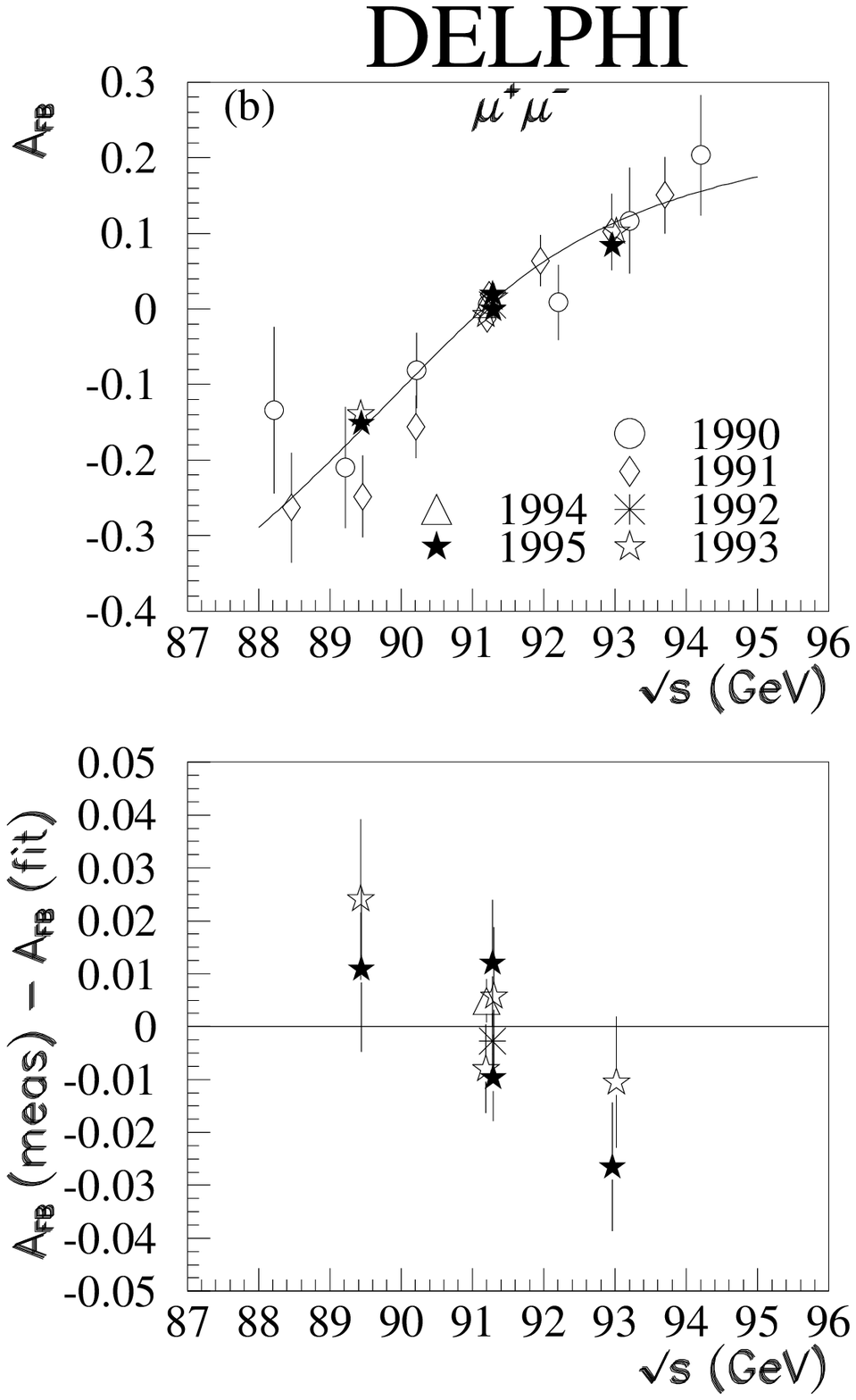}
    \caption[]{\label{fig:afb}
              The results of the measurement of ${\cal A}_{FB}$
              in $\EEMM$ by the
              DELPHI collaboration.
            }
\end{figure}

The measurement of ${\cal A}_{FB}$ is done for the leptonic final states,
for b- and c-quarks separately and averaged over all quarks.
As an example, ${\cal A}_{FB}$
measured in $\EEMM$ as a function of $\sqrt s$~ by 
the DELPHI experiment~\cite{delphiz}
is shown in  Figure~\ref{fig:afb}. 
The points are the data
at several $\rts$ and the curve is the result of a fit using 
ZFITTER. In the lower plot the difference between the data and 
the fitted curve is also shown.

More information about the
couplings is obtained by the 
measurement of the polarisation of the fermions
in the final state. 
Denoting with $h_f$ the helicity of the
final state fermion one obtains
for unpolarised beams
:    
\begin{eqnarray*}
{\cal P} &=& \frac{\sigma(h_f = +1)-\sigma(h_f = -1)}{\sigma(h_f = +1)+
           \sigma(h_f = -1)}  \\
         &=& -\frac{2v_f a_f}{v_f^2+a_f^2} = -{\cal A}_f
\end{eqnarray*}
and
\begin{eqnarray*}
{\cal P}_{FB} = 
               -\frac{4v_e a_e}{v_e^2+a_e^2} = -2{\cal A}_e.
\end{eqnarray*}
${\cal P}_{FB}$ denotes the  forward-backward polarisation asymmetry.
As can be seen, the measurement of 
the polarisation of the final state fermion 
allows to determine, separately
for the electron and the final state fermion, the product
of vector and axial-vector couplings.

\begin{figure}[htb]
  \begin{center}
    \includegraphics[width=0.5\textwidth]{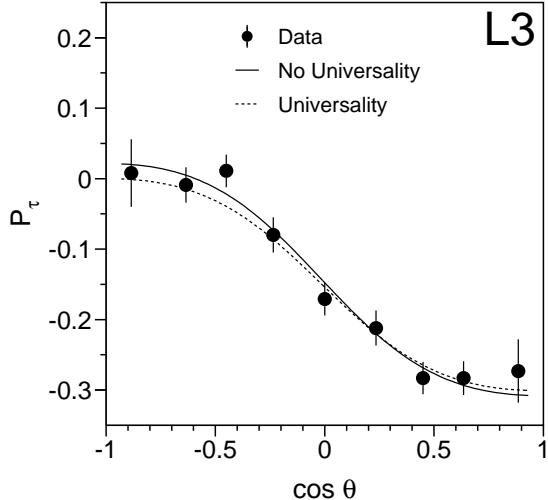}
    \caption[]{\label{fig:l3taup}
            The polarisation of the $\tau$ in  $\EETT$
            as function of $\cos \theta$ measured by the L3 
            experiment. 
            }
  \end{center}
\end{figure}

\begin{figure}[htb]
    \hspace{-0.3cm} 
    \includegraphics[width=0.53\textwidth]{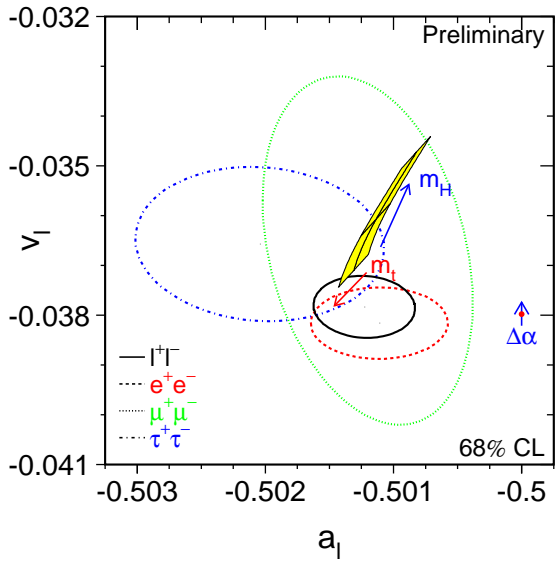}
    \caption[]{\label{fig:lepcoup}
              Lepton couplings determined by the LEP experiments
              and the measurement of ${\cal A}_{LR}$ from SLD~\cite{lepew}.
            }
\end{figure}

In the experiments the polarisations
are only accessible in
the process $\EETT$, because the $\tau$ lepton has a very short lifetime
and decays inside the detector. From the analysis of the $\tau$
decay products, assuming (V-A) structure of the charged current, 
the polarisation of the $\tau$ is measured.
Also these measurements were done by all LEP collaborations
and as an example the result of the L3 experiment~\cite{l3tau}
is shown in Figure~\ref{fig:l3taup}.
Taking the measurements of cross sections, forward-backward asymmetries and 
the measurement of the $\tau$ polarisation together,
and using also ${\cal A}_{LR}$\footnote{${\cal A}_{LR} \sim {\cal A}_{e}$ 
is the cross section asymmetry for left-handed
and right-handed electrons in $\EE$ annihilations
measured with the polarised electron beam at SLAC.} 
from SLD~\cite{sld},
the vector and axial-vector couplings
of the leptons are determined. This is shown in Figure~\ref{fig:lepcoup}
first for 
each lepton flavour separately.
As can be seen, the couplings of electrons, muons and
taus are in agreement, supporting the fundamental assumption of
lepton universality.

\stepcounter{section}
\section*{\large \arabic{section}.\ \large Electroweak radiative corrections
}

\begin{figure}[htb]
  \begin{center}
    \includegraphics[width=0.35\textwidth]{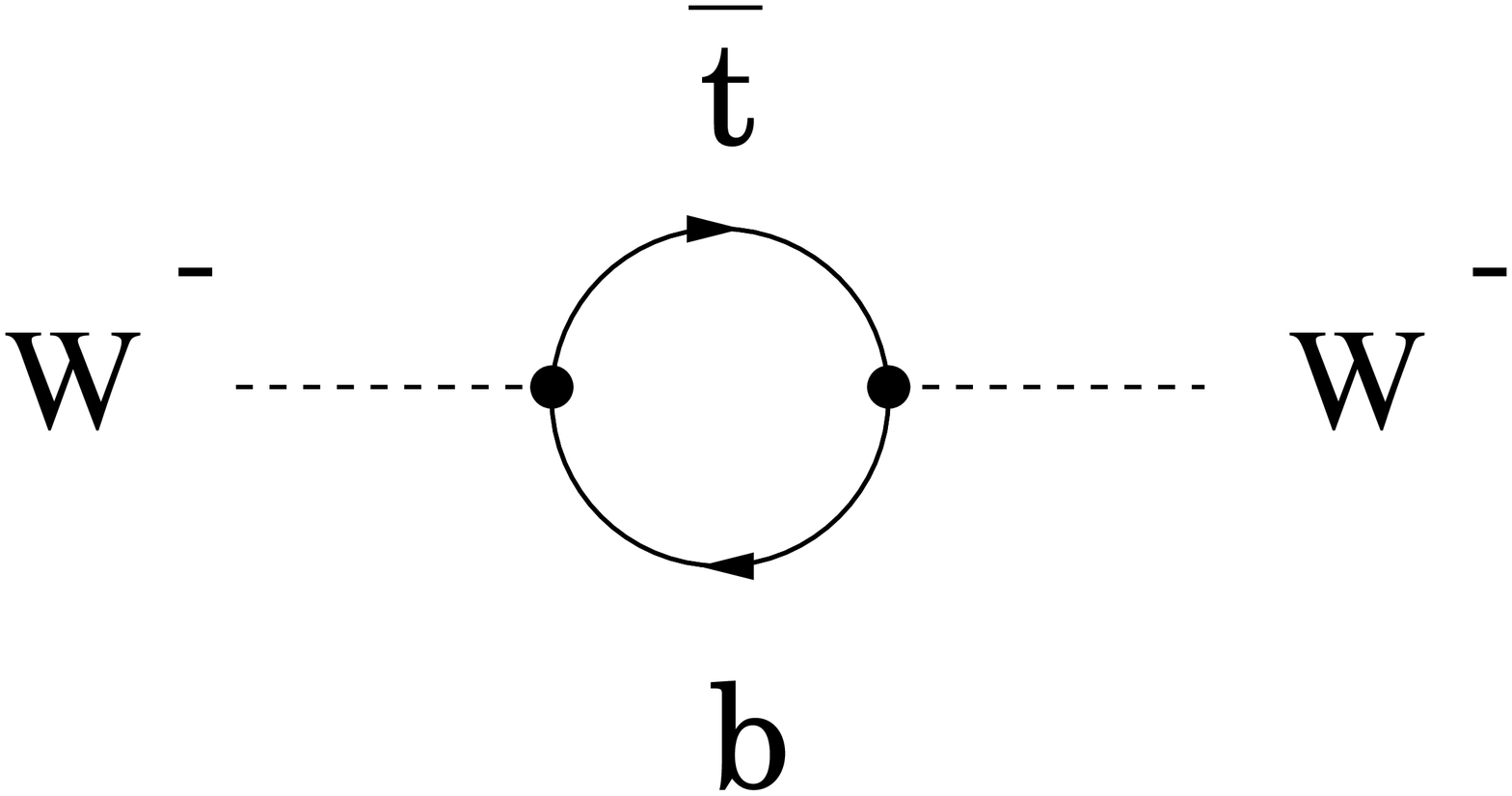}
    \phoo \\
    \VS{0.2}
    \phoo  
    \includegraphics[width=0.22\textwidth]{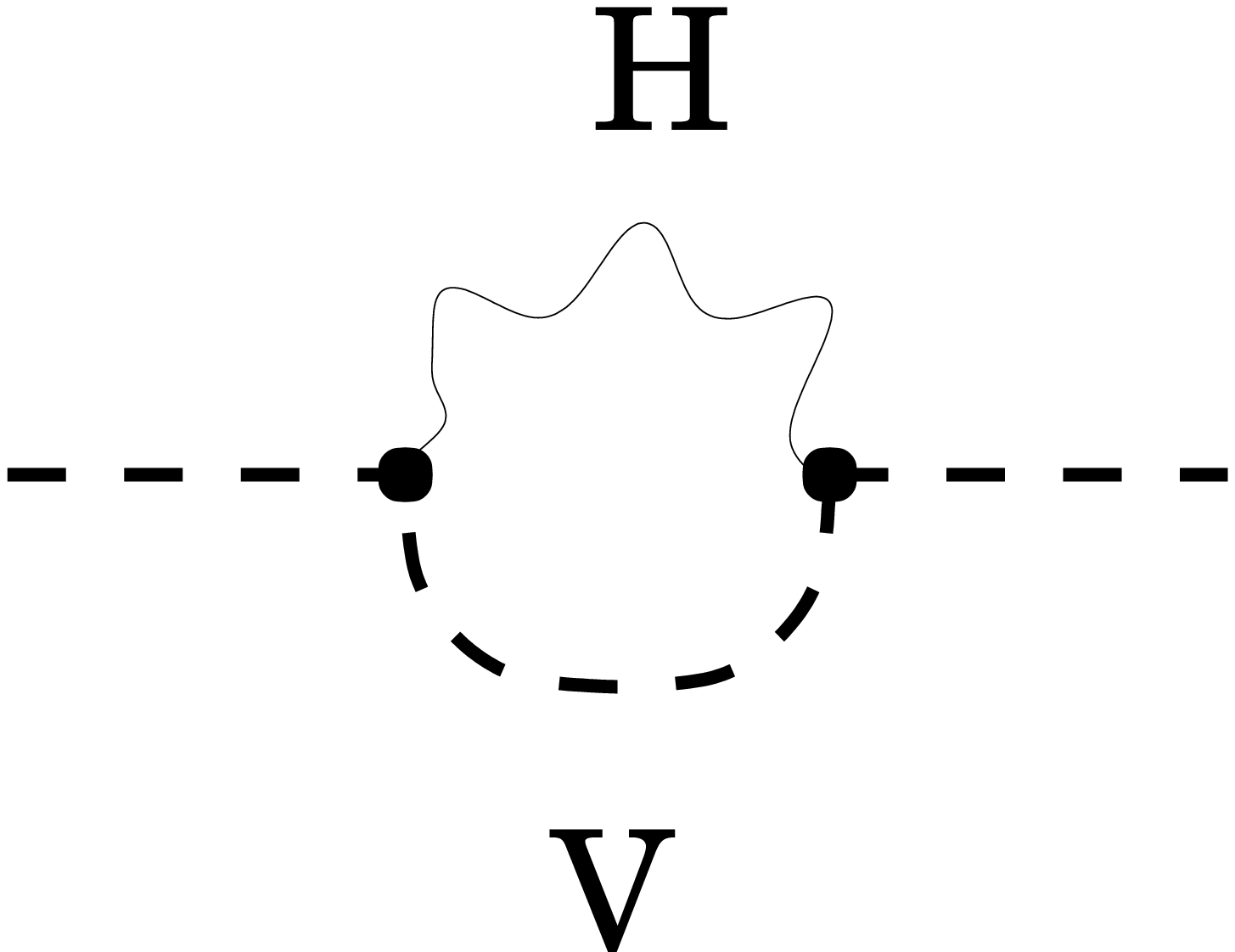}
    \caption[]{\label{fig:weaklo}
              Lowest order weak correction diagram
              containing fermion loops (top) or the Higgs boson (bottom).
              V is either Z or W.  
            }
  \end{center}
\end{figure}

Before interpreting the results on the couplings the phenomenon
of electroweak loop corrections should be discussed.
From QED the vacuum polarisation, described
by fermion loops, 
is well known.
Such corrections depend on the mass of the fermions
and can easily calculated for leptons for which masses are precisely 
known.
For quarks they are calculated from the low energy
cross sections $\EE \rightarrow hadrons$
via dispersion relations~\cite{eidel}.
The vacuum polarisation leeds to an increase of the 
effective electromagnetic coupling with $s$
and at an energy corresponding to the Z mass this results in 
$\alpha (\MZ^2) = \frac{\alpha(0)}{1-\Delta \alpha(\MZ^2)} = 1./128.945$.
In addition
there are weak loop corrections, for which examples are shown in 
Figure~\ref{fig:weaklo}.
These corrections are quantified as $\Delta r_W$ in the relation
\begin{eqnarray*}
\gf = \frac{\pi \alpha}{\sqrt 2 {\MW}^2 s_W^2} \frac{1}{1-
\Delta \alpha(s) - \Delta r_W } ,
\end{eqnarray*}
where $\MW$ is the mass of the W boson. 
The upper diagram
of Figure~\ref{fig:weaklo}
 results in contributions 
to $\Delta r_W$ proportional to $\simeq \MTOP^2/\MW^2$,
where $\MTOP$ is the mass of the top-quark.
In addition, there are
bosonic loops containing the Higgs boson. They
are proportional to $\simeq ln~\MH^2/\MW^2$~\cite{rc}. 

In order to compare the measured couplings to the predictions
of the theory, the improved Born approximation~\cite{imbo} is used,
in which the electroweak loop corrections
are absorbed in effective couplings:
\begin{eqnarray*}
\tilde{v}_f &=&  \sqrt{\rho}(I_3-2 \kappa s_W^2) \\
\tilde{a}_f &=&  \sqrt{\rho} I_3.
\end{eqnarray*}
Here
$\rho$~\cite{velt}
is the ratio of neutral to charged current couplings,
$\rho =c_w^2\MZ^2/\MW^2$.
At Born level holds $\rho$ = 1. Weak loop corrections
lead to $\rho$ = 1 + $ \Delta \rho$, with 
$ \Delta \rho = -\frac{s_W^2}{c_W^2} \Delta r_W - \Delta r_{rem}$.
The quantity $\Delta r_{rem}$ containes non-leading contributions.  
In addition, weak loop corrections, similar in size,  
contribute to $\kappa$.

The exploitation of these 
relations was very successful at LEP before the discovery 
of the top-quark. Using the precision measurements
around the Z, the $\simeq \MTOP^2/\MW^2$
dependence of
the weak loop contributions allowed
to predict the top-quark mass in 1993 
to be $\MTOP = 166^{+17+19}_{-19-22}$~\cite{ewold},
in excellent agreement with the measurement after the 
discovery 1995~\cite{topdisc}.

As shown in Figure~\ref{fig:lepcoup},
lepton universality is supported by the measurements.
Hence the results from the leptonic final states
are combined to derive universal lepton
couplings.
The result, also shown in  Figure~\ref{fig:lepcoup},
leads to
a value of $a_l$
clearly differerent from the prediction of the SM of -0.5, 
pointing to significant contributions from weak corrections.
Interpreting the measurements in
the improved Born approximation
 the value of $\rho$ is determined to be 
1.0050 $\pm$ 0.0010~\cite{lepew},
different from unity by 5 standard deviations.

\begin{figure}[htb]
\HS{-0.5}
    \includegraphics[width=0.56\textwidth]{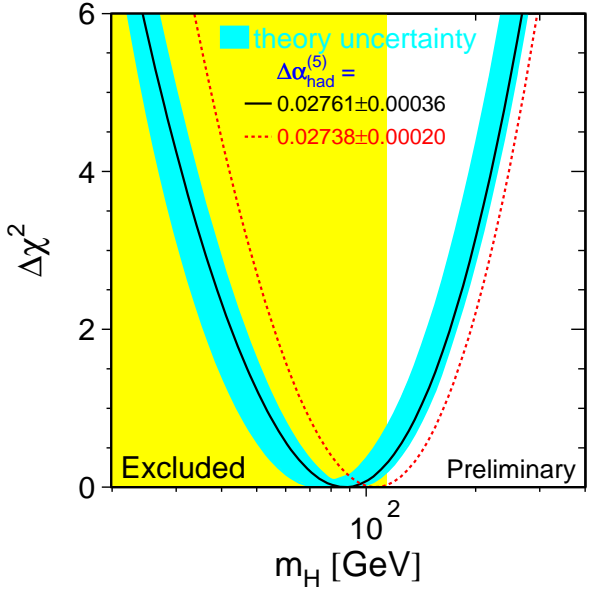}
    \caption[]{\label{fig:higgsew}
              The $\chi^2$ of a fit to all electroweak measurements
              with the Higgs boson mass as free parameter~\cite{lepew}.  
            }
\end{figure}

Using, in addition to LEP and SLD data, also results on $\MW$~\cite{mwtev}
and $\MTOP$~\cite{mttev} from the Tevatron and on atomic parity violation
\cite{pato}
a fit in the framework of the SM is performed with
$\MH$ as free parameter.
In  Figure~\ref{fig:higgsew} the dependence of the
$\chi^2$ of the fit on the mass of the Higgs boson
is shown. 
The value
for the Higgs boson mass obtained 
is $\MH = 88^{+53}_{-35} \GeV$, or, at 95\% C. L., $\MH <$ 196 \GeV. 
Also shown in the Figure is the light grey area 
excluded from direct searches of the Higgs boson. 
Apparently, the electroweak 
precision measurements favour a light Higgs boson which might be just
a bit above the
current lower mass limit from direct searches. 

\stepcounter{section}
\section*{\large \arabic{section}.\ \large $\EE$ annihilations above the Z}

After 1995 the beam energy
of the LEP accelerator was enlarged in several steps 
up to 104.5 \GeV, reached in the year 2000. 
In addition to the annihilation into fermion-antifermion pairs
we expect several new channels kinematically
allowed and subject of new kind of studies. These are:
\begin{itemize}
\item{$\EE \rightarrow \WW$}: The pairwise production of
W bosons allows the measurement of their mass, and, for the first time under
very clean conditions, the study of triple gauge boson couplings.
\item{$\EE \rightarrow \ZZ$}: In the SM this process
is mediated by virtual electron exchange in the t-channel.
The cross section is sensitive to anomalous
gauge boson couplings.
\item{$\EEZH$}: This process is expected in the Standard Model.
But since the mass of the Higgs boson 
is unknown, we do not know, whether it
is accessible at LEP energies.
\item{$\EE \rightarrow$ new particles}: If there is physics beyond the 
Standard Model it can manifest itself \eg~via pair production
of supersymmetric particles.      
\end{itemize}

\stepcounter{subsection}

\subsection*{\normalsize \thesubsection.\ $\EFFG$}
\begin{figure}[htb]
  \begin{center}
    \includegraphics[width=0.5\textwidth,height=0.5\textheight]{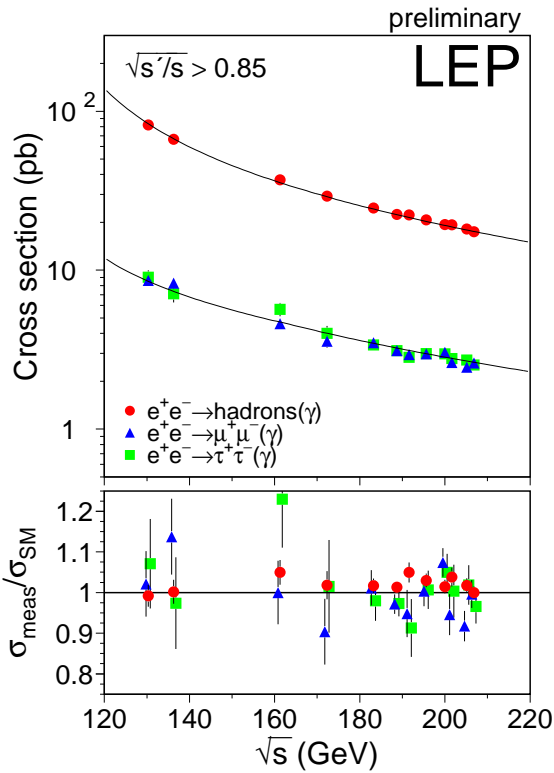}
    \caption[]{\label{fig:csff}
              The cross section
              measurements for the processes
              $\EE \ra hadrons$ and $\EE \rightarrow \LL$ ($\ell = $e, $\mu$),
              at high energies combined for all LEP experiments. 
            }
  \end{center}
\end{figure}

This process is mediated by virtual photon or Z exchange. 
Some fraction of the events is characterised
by the radiation of a high energy photon in the 
initial state, reducing the
effective annihilation energy to the mass of the Z.
Of interest are, of course, the events 
where the annihilation happens with the full energy.
The cross sections measured at high energy
for $\EE \ra hadrons$, $\EEMM$ and 
$\EETT$ are shown in  Figure~\ref{fig:csff}~\cite{lepff}
for the fraction
with full annihilation energy. 
The full lines
are the expectations from the SM. As can be seen, the SM describes perfectly
the dependence of the annihilation cross section on $\rts$.
These results are used
to derive limits on new physics scales, \eg~the scale
of contact interactions~\cite{cint}. Such interactions are excluded
up to several TeV~\cite{lepff} for any helicity
structure of the amplitudes. Also stringent limits on 
couplings and masses of leptoquarks
or additional heavy gauge bosons are derived from these measurements.
For example, a sequential heavy gauge boson 
is excluded up to a mass of 1.89 TeV.  

\stepcounter{subsection}
\subsection*{\normalsize \thesubsection.\ W production and decay}

Both single and pair production of W bosons are new
physics topics at high energies. 
From $\EEWW$ and $\EEWEN$ 
the mass of the W, $\MW$, the width of the W,
and the couplings
between gauge bosons, WW$\gamma$ and WWZ, 
are measured.

The event topologies we expect
for a $\WW$ final state are four 
jets when both W decay into quarks, $\rm W \rightarrow \rm q \bar{\rm{q}}'$,
two jets and one charged lepton 
when one W decays into quarks and 
the other into leptons, $\rm W \rightarrow \ell \nu$,
or two charged 
leptons if both W decay leptonically. 

\begin{figure}[htb]
    \includegraphics[width=0.5\textwidth]{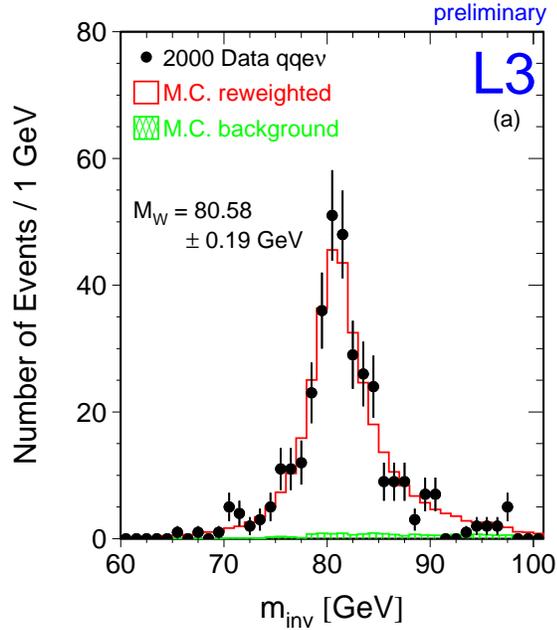}
    \caption[]{\label{fig:wmass}
              The measurement of the W mass by the L3 experiment in the 
              final state $\rm q\bar{\rm{q}}'\rm{e} \nu$.
            }
\end{figure}

\begin{figure}[htb]
  \begin{center}
    \includegraphics[width=0.55\textwidth]{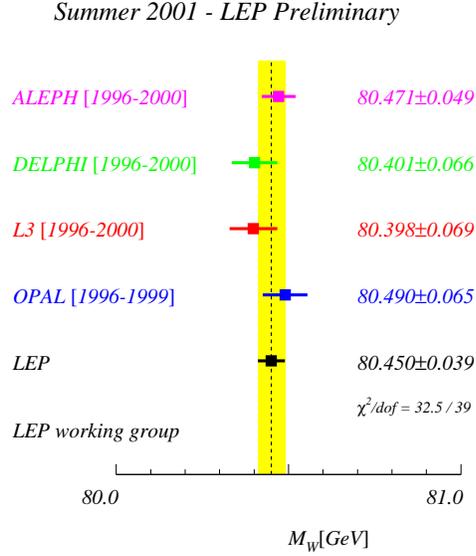}
    \caption[]{\label{fig:wmasscomp}
              The measurement of the W mass by the
              LEP collaborations and their combination~\cite{lepew}.
            }
  \end{center}
\end{figure}

The mass of the W is determined from the direct reconstruction of the
invariant mass of the two jets or the two leptons. 
In the latter case the momentum of the 
neutrino is taken as the missing momentum of the event.    
An example for a such reconstructed mass distribution is shown in
Figure~\ref{fig:wmass}
for the final state $\rm q\bar{\rm{q}}'\rm{e} \nu$~\cite{l3ww}.
 The distribution shows a clear,
nearly background free, resonance shape.
The measurement is done using all channels and the results 
of the LEP experiments are given in Figure~\ref{fig:wmasscomp}. 
Combining them $\MW = 80.450 \pm 0.039$ \GeV is obtained.

\begin{figure}[htb]
  \begin{center}
    \includegraphics[width=0.5\textwidth, height=0.2\textwidth]{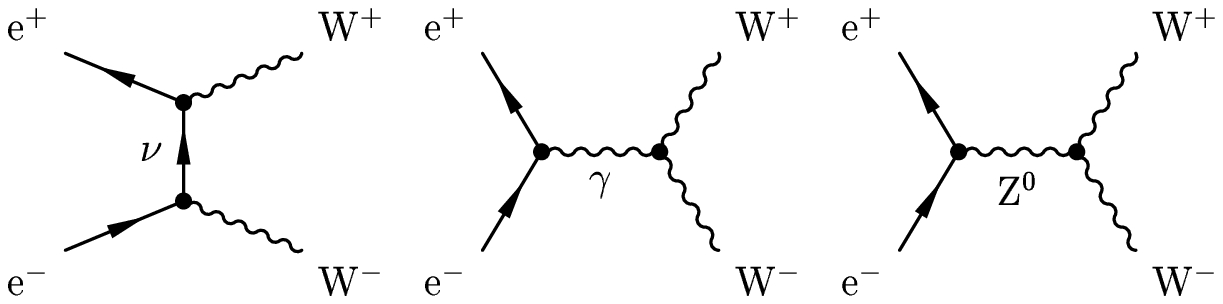}
    \caption[]{\label{fig:wwprod}
              The processes contributing to $\EEWW$.
            }
  \end{center}
\end{figure}

The process $\EEWW$ is of particular interest
because triple gauge boson couplings~\cite{hagiwa},
predicted in non-abelian gauge theories, 
can be measured.
In the SM three diagrams as shown in Figure~\ref{fig:wwprod}
contribute to $\EE \rightarrow W^+W^-$.
 Of particular
interest are the latter two, because only they contain triple gauge boson 
couplings.
Furthermore, the cross section obtained from each diagram separately
is divergent with growing $\sqrt s$; only the sum of the three diagrams
results in a finite cross section.

\begin{figure}[htb]
\HS{-0.5}
    \includegraphics[width=0.55\textwidth]{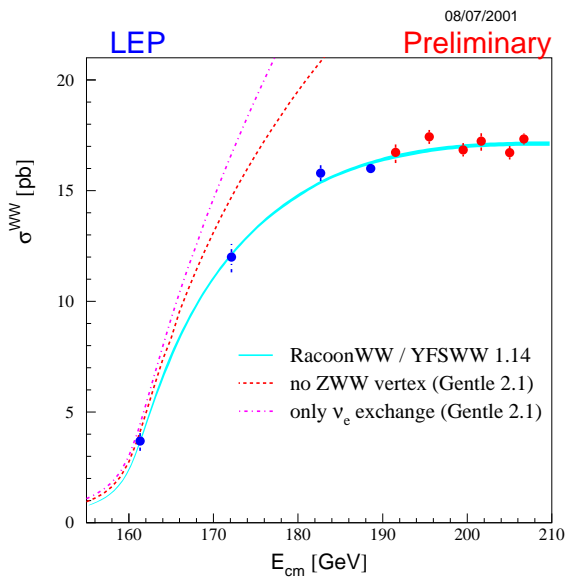}
    \caption[]{\label{fig:wwprodcs}
              The cross section measurements for $\EEWW$
              and the predictions of the several subprocesses~\cite{lepew}.
            }
\end{figure}

The measurements of the cross sections and the predicted behaviour
for only neutrino exchange, neutrino plus photon exchange and 
and the sum of all contributions is shown in Figure~\ref{fig:wwprodcs}.
The data clearly favour the SM prediction.
The $\gamma$WW and ZWW vertices
are
described by $2 \times 7$ independent form factors~\cite{hagiwa}.
Assuming CP and P conservation 
and electromagnetic gauge invariance,
five independent couplings are left, $g^Z_1, \kappa_\gamma,
\kappa_Z, \lambda_\gamma$ and $\lambda_Z$, with values predicted 
in the SM of 1, 1, 1, 0 and 0. 
The first coupling is $g^Z_1 = \frac{g_{ZWW}}{g^{SM}_{ZWW}}$, with 
$g^{SM}_{ZWW} = g'c_W$ and the other   
quantites are related to
static properties of the W,~\eg~the magnetic dipole moment
$\mu_W=\frac{e}{2\MW}(1+\kappa_\gamma+\lambda_\gamma)$
or the electromagnetic quadrupole moment
$Q_W=\frac{-e}{\MW^2}(\kappa_\gamma-\lambda_\gamma)$. 
Assuming $SU(2)$ gauge invariance, the following relations hold:
\begin{eqnarray*}
\Delta \kappa_Z &=& \Delta g_1^Z \Delta \kappa_\gamma \tan^2 \Theta_W  \\
\lambda_Z &=& \lambda_\gamma,
\end{eqnarray*}  
where $\Delta\kappa$ and $\Delta g_1^Z$ denote the 
deviation of the quantities from the SM value. 
Exploiting the total cross section measurements,
the production and decay angular distributions of the W bosons
and using also the measurements from single W production
the three
left parameters are determined. Fits, in which one parameter is
free and the other two are set to their SM values,
yield~\cite{lepww}:
\begin{eqnarray*}
\Delta g_1^Z &=& -0.025 \pm 0.026,   \\
\Delta \kappa_\gamma &=& -0.002 \pm 0.067, \\
\lambda_\gamma &=& -0.036 \pm 0.028.
\end{eqnarray*}
 
The results of two-dimensional fits give similar results, being in
agreement with the SM expectations.

\stepcounter{section}
\section*{\large \arabic{section}.\  Searches} 

\stepcounter{subsection}
\subsection*{\normalsize \thesubsection.\ Higgs boson phenomenology}

\begin{figure}[htb]
  \begin{center}
    \includegraphics[width=0.45\textwidth]{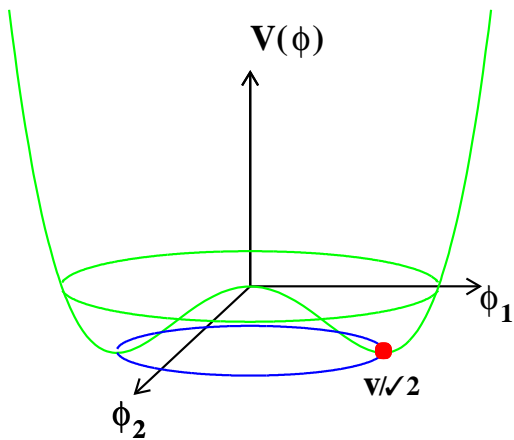}
    \caption[]{\label{fig:hipot}
              The shape of the potential of the Lagrangian
              describing the scalar $SU(2)$ doublet. 
            }
  \end{center}
\end{figure}

The Higgs mechanism is introduced
to allow particles to acquire mass keeping the gauge invariance
of the SM Lagrangian. In its minimal version a
$SU(2)$ doublet 
of complex fields is introduced:
\begin{eqnarray*}
\Phi ={\Phi_1(x) \choose \Phi_2(x)}.
\end{eqnarray*}
The simplest Lagrangian reads
\begin{eqnarray*}
{\cal{L}}_H &=& \partial_\nu \Phi^{\dagger}\partial^\nu \Phi -{\cal{V}}(\Phi),
\end{eqnarray*}
with the following choice of the potential
\begin{eqnarray*}
{\cal{V}}(\Phi) &=&-\mu^2 \Phi^{\dagger} \Phi + \lambda (\Phi^{\dagger} \Phi)^2.
\end{eqnarray*}    
The state with minimal energy corresponds to the minimum of the
potential. Defining
\begin{eqnarray*}
\frac{\upsilon}{\sqrt 2} = \sqrt {\Phi^{\dagger} \Phi},
\end{eqnarray*}
the potential reads:
\begin{eqnarray*}
{\cal{V}}(\Phi) = -\frac{1}{2}\mu^2\upsilon^2+\frac{1}{4}\lambda\upsilon^4.
\end{eqnarray*}
The minimum of the potential is reached at 
$\upsilon = \upsilon_0 = \sqrt {\mu^2/\lambda}$.
This is illustrated in 
Figure~\ref{fig:hipot}. 
In the plane $\Phi_1$ vs. $\Phi_2$ the minimum of the potential 
is continously degenerated. The choice
\begin{eqnarray*}
\Phi_0 ={0 \choose \frac{1}{\sqrt 2}\upsilon_0},
\end{eqnarray*}
leads to a ground state, which is,
in contrary to the potential, not 
invariant under $SU(2)$ transformations,
a feature denoted as spontanous symmetry breaking. 
After 
redefinition of the scalar field,
$H(x) = \upsilon(x) - \upsilon_0$,
we inspect the covariant terms of the
Lagrangian
\begin{eqnarray*}
{\cal{L}}_H &=& {\cal{D}}_\nu \Phi^{\dagger} {\cal{D}}^\nu\Phi -{\cal{V}}(\Phi).
\end{eqnarray*}
It results into expressions
\begin{eqnarray*}
\frac{1}{4}g'^2 \upsilon_0^2(W^-_{\lambda}W^{+{\lambda}}) 
\end{eqnarray*}
and
\begin{eqnarray*}
\frac{1}{8} (g^2+g'^2) \upsilon_0^2 Z_{\lambda}Z^{\lambda},
\end{eqnarray*}
which we interprete as mass terms for the W and Z bosons:
\begin{eqnarray*}
\MW^2 &=& \frac{1}{4}g'^2\upsilon_0^2  \\
\MZ^2 &=& \frac{1}{4}(g^2+g'^2)\upsilon_0^2.
\end{eqnarray*}
The interaction between
gauge bosons and the Higgs boson is described by the terms:
\begin{eqnarray*}
\frac{1}{2} g'^2 \upsilon_0(W^-_{\lambda}W^{+{\lambda}})H 
\end{eqnarray*}
and
\begin{eqnarray*}
\frac{1}{4} (g^2+g'^2) \upsilon_0 (Z_{\lambda}Z^{\lambda})H.
\end{eqnarray*}
In addition, there appear the terms
$2 \lambda \upsilon_0^2H^2$, $\lambda \upsilon_0 H^3$
and $\frac{1}{4} \lambda H^4$. 
The first is the mass term of the physical state of the
scalar field, the Higgs boson, with 
$\MH^2 = 2 \lambda \upsilon_0^2$.
The other two correspond to self couplings 
of the Higgs field.
The parameter $\upsilon_0$, the value of the potential in the
ground state, is known from the measurement of $\gf$
to be $\upsilon_0 \simeq$ 246 \GeV. 
The parameter $\lambda$, determining the shape
of the potential, appears in the mass term for the Higgs boson
and in addition in the self couplings.

The discovery of the Higgs boson and the measurement
of its mass will be the 'key-stone' of the SM.
The measurement of the self couplings would confirm the 
shape of the potential as shown in 
Figure~\ref{fig:hipot}.
Also terms describing quartic interactions between gauge
bosons and Higgs bosons, not discussed here, appear.

The masses of the fermions are obtained from
separately introduced Yukawa terms,
\eg~for electrons, as
\begin{eqnarray*}
{\cal{L}}_Y &=& \sqrt{2} \HEE \rm{\bar{e}_R} \Phi^{\dagger}{\nu_e \choose \rm{e_L}} +h.c. \\            &=& \HEE \upsilon_0(\rm{\bar{e}_L}\rm{e_R}+
\rm{\bar{e}_R}\rm{e_L})
\end{eqnarray*}
with
$m_e = \HEE \upsilon_0$.

\stepcounter{subsection}
\subsection*{\normalsize \thesubsection.\ Higgs boson search}

\begin{figure}[htb]
  \begin{center}
    \includegraphics[width=0.5\textwidth]{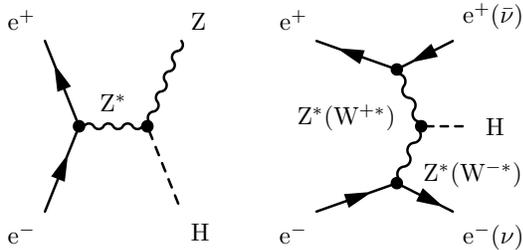}
    \caption[]{\label{fig:hiprod}
              The production mechanisms of the Higgs boson
              at LEP: Higgs-strahlung(left) and fusion (right).
            }
  \end{center}
\end{figure}

\begin{figure}[htb]
  \begin{center}
    \includegraphics[width=0.53\textwidth]{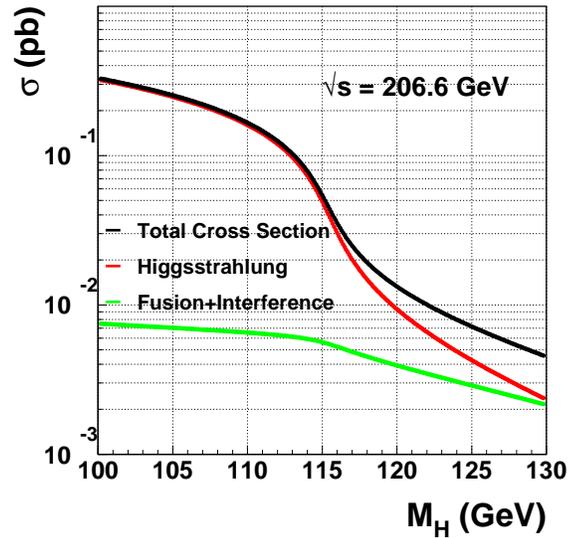}
    \caption[]{\label{fig:hiprodcs}
              The cross sections  of Higgs-strahlung and fusion
               processes
              at a centre-of-mass energy of $\simeq$ 207 \GeV~
              as function of $\MH$.
            }
  \end{center}
\end{figure}

\begin{figure}[htb]
  \begin{center}
    \includegraphics[width=0.48\textwidth]{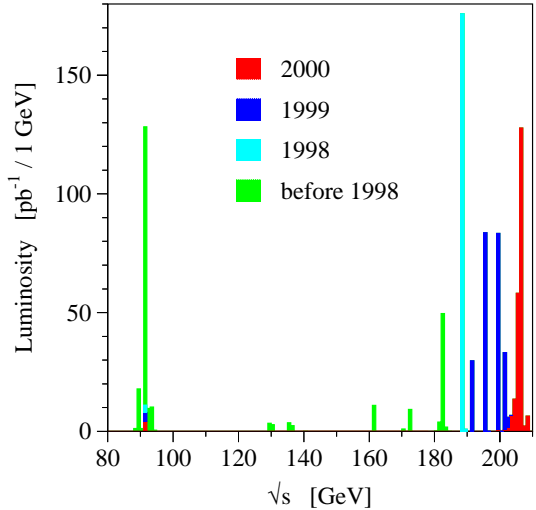}
    \caption[]{\label{fig:leplumi}
              The integrated luminosities collected
              by all LEP experiments
              as function of $\rts$. Of importance for the
              Higgs boson search at large $\MH$ are the data 
              with at the highest energies.
            }
  \end{center}
\end{figure}

The production of the Higgs boson in $\EE$ annihilations
proceeds at LEP energies dominantly via the Higgs-strahlung, 
illustrated in Figure~\ref{fig:hiprod}.
Higgs-strahlung reaches at a given centre-of-mass energy 
the kinematic limit if $\MH$+$\MZ$ approaches $\rts$.
The fusion process contributes
below the kinematic threshold of the Higgs-strahlung only with a very small 
fraction to the cross section. This can be seen from 
Figure~\ref{fig:hiprodcs}.
Since the integrated luminosity
of LEP at the highest energies, shown in  Figure~\ref{fig:leplumi},
is about 100 pb$^{-1}$, only the Higgs-strahlung is of practical
importance. 

\begin{figure}[htb]
  \begin{center}
    \includegraphics[width=0.55\textwidth]{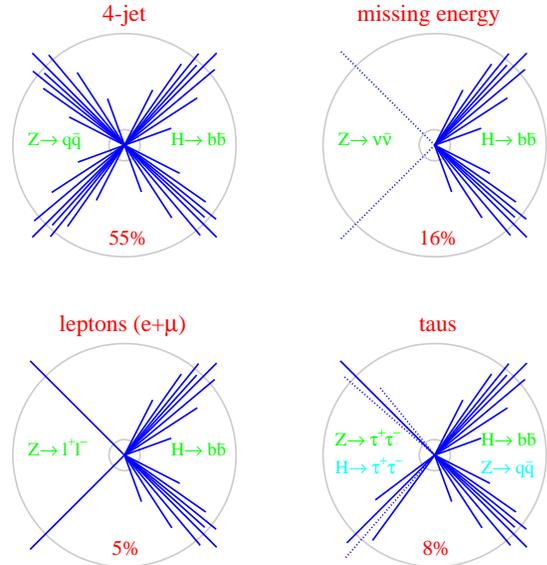}
    \caption[]{\label{fig:hitopo}
              The event topologies for the dominant Higgs boson
              and Z decays
            }
  \end{center}
\end{figure}

\begin{figure}[htb]
  \begin{center}
    \includegraphics[width=0.5\textwidth]{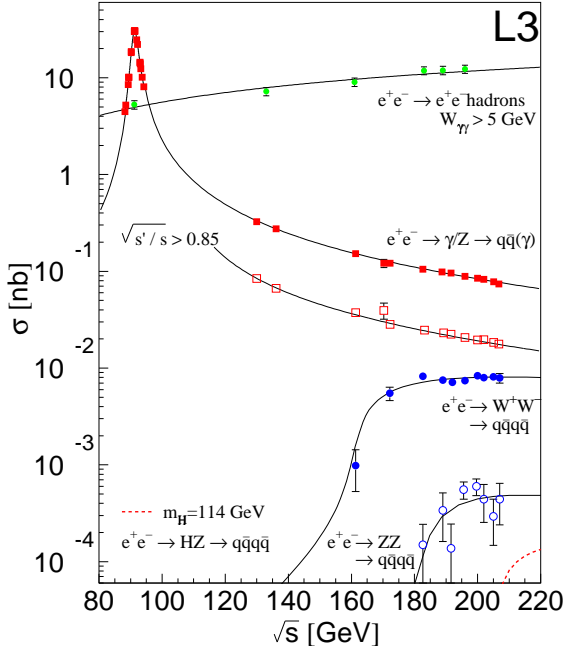}
    \caption[]{\label{fig:backg}
            The cross sections of the main background processes
            in the Higgs boson search. For illustration, the expected
            cross section for a Higgs boson with $\MH$ = 114 \GeV~
            is indicated as dashed line in the lower right edge. 
            } 
  \end{center}
\end{figure}

The decay of the Higgs boson into a fermion-antifermion pair
is determined by the Yukawa coupling, $\HFF$, which is proportional 
to the fermion mass. Hence the Higgs boson is expected 
to decay dominantly into the heaviest fermion-antifermion pair
kinematically allowed. In the mass range accessible at LEP 
the decay into b-quarks is dominant, also the decay into 
a $\TT$ is of importance.
Taking into account the decay channels of the Z, the following
final states must be considered:
$\zhqqbb$,
$\zhnnbb$,
$\zhllbb$,
and $\zhqqtt$.
They result into the event
 topologies
shown in Figure~\ref{fig:hitopo}.
However, $\EE$ annihilation processes of large cross section,
as shown in Figure~\ref{fig:backg}, 
result in final states, which can fake a Higgs boson signal.
In order to suppress the background and to isolate events originating from 
Higgs boson production dedicated techniques are developed.
Of particular importance is the detection of b-quarks
in jets assumed to stem from the Higgs boson.
After hadronisation, b-quarks lead to 
B-hadrons in the jets with a lifetime
$\tau_{B}$ of about 1.5 picoseconds. 
This lifetime is sufficient for a decay length
of several mm, leading to secondary vertices. 
The reconstruction of the  secondary vertices is achieved
with silicon microvertex detectors.
In addition, the presence of a Z is exploited requiring 
the invariant mass of the two jets not assigned to the Higgs
boson or the two leptons to be equal to $\MZ$.
Other quantities used to discriminate signal from background are event 
shape variables,
characterising the kinematics of
the jets, or boson production angles.

\begin{figure}[htb]
  \begin{center}
    \includegraphics[width=0.55\textwidth]{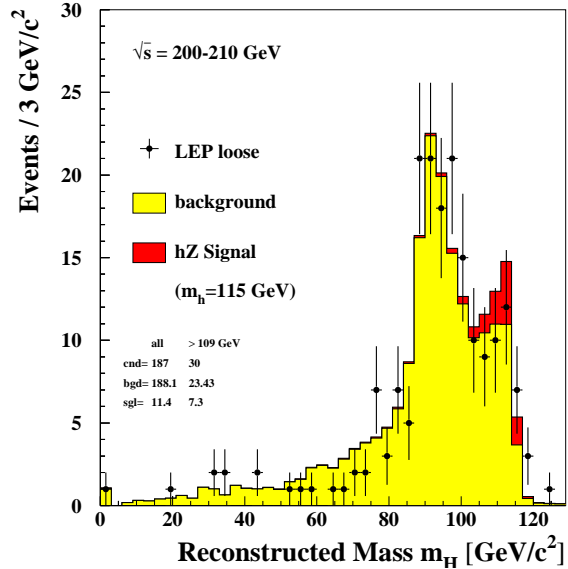}
    \caption[]{\label{fig:signal}
             The invariant mass of the two jets assigned
             to the Higgs boson for data, background and signal
             events requiring S/B$>$ 0.5.  
             The mass of the Higgs boson is set to $\MH = 115 \GeV$
             \cite{higgswg}. 
            } 
  \end{center}
\end{figure}

\begin{figure}[htb]
  \begin{center}
    \includegraphics[width=0.53\textwidth, height=0.55\textwidth]{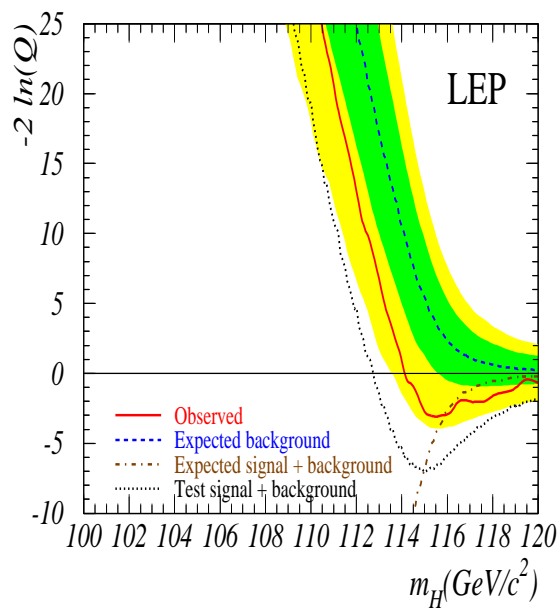}
    \caption[]{\label{fig:statana}
             The quantity $-2\ln Q(\MH)$ obtained from the Higgs 
             boson search results of all LEP experiments combined
             as function of $\MH$. The dark and light grey areas
             indicate the 1$\sigma$ and 2$\sigma$ significances. 
            } 
  \end{center}
\end{figure}

At the end of the analysis chain usually a
few events remain which 
are in their features similar to the expected signal.
The fraction of expected
background 
and  signal events is estimated from
MC simulations.
In order to gain sensitivity, the results of
the four LEP experiments are combined~\cite{higgswg}.
For that purpose to each selected event a weight
is assigned which is obtained in the following way:
The event characteristics important 
for the detection
of a Higgs boson signal are used to construct a 
discriminant. For each channel the selected events are distributed
as function of the discriminant.
Events from all background processes
and from the expected signal, which survive the analysis,
are distributed in the same discriminant.
For each bin of the distribution the ratio
signal/background ($S/B$) is calculated, 
and this quantity is used to calculate
the weight of the data events in the same bin of the distribution.
As an example, Figure~\ref{fig:signal} shows
the distribution of the invariant mass of the jets assigned to 
the Higgs boson   
for data, MC background and an expected signal with $\MH = 115 \GeV$
for an event sample with $S/B >$ 0.5\footnote{This sample 
is obtained after application of cuts on the most sensitive quantities
with no or only litle dependence on $\MH$.}. 
As can be seen data 
follow the distribution expected
from the background. 
Finally, dedicated statistical analysis
is applied to test the compatibility
of the observed data with the expectation for
only background and signal plus background hypotheses.
The crucial quantity is the 
likelihood ratio
\begin{eqnarray*}
Q({\MH}) = \frac{{\cal{L}}(S+B)}{{\cal{L}}(B)} .
\end{eqnarray*}
The logarithm of Q reads:  
\begin{eqnarray*}
-2\ln Q(\MH)= ~~~~~~~~~~~~~~~~~ \\
2 s_{tot} -2\sum_{i}N_i(1+\frac{S_i}{B_i}). 
\end{eqnarray*}
The index i runs over all bins
of the discriminant distribution, $s_{tot}$ is the total 
expected signal rate 
and $N_i$ the number of data events in the bin.
The quantity $-2\ln Q(\MH)$ is shown in Figure~\ref{fig:statana}
as function of $\MH$.
The dashed line shows the expectation for the background, 
the dashed-dotted line for a signal plus background
and the full line the result for the data.
The dotted line is expected if a Higgs boson with
$\MH = 115 \GeV$ would be present.
There is some excess in data around $\MH = 115 \GeV$
with a statistical significance of 2.1 $\sigma$.
This effect is
due to an excess in the four-jet topology and
these events originate mainly
from the ALEPH experiment. 

Since no significant signal is found the data are used to set a mass 
limit for the Higgs boson, exploiting the strong dependence
of the Higgs boson production cross section on $\MH$
shown in Figure~\ref{fig:hiprodcs}.
Taking again the results from all four experiments
this lower mass limit is $\MH >$ 114.1 $\GeV$
(95 \% C.L.).

\stepcounter{subsection}
\subsection*{\normalsize \thesubsection.\ Supersymmetry}

Supersymmetry is a tempting theoretical concept 
~\cite{susy}, which avoids
the so called 'hierarchy problem'~\cite{hiera} and includes in a 
natural way gravity. It predicts supermultiplets which 
contain particles of different spins; $\eg$ the SM fermions
of a $SU(2)$ doublet or singlet
are supplemented by scalar 'sfermions' and the bosons
by spin $\frac{1}{2}$ particles denoted
vinos, zinos, higgsinos and gluinos. Gravitons 
with spin 2 are grouped with gravitinos of spin $\frac{3}{2}$.  
Unbroken supersymmetry would add, apart of many new particles,
only one free parameter to the SM, called $\tan \beta
= \frac{\upsilon_1}{\upsilon_2}$, where $\upsilon_1$ and $\upsilon_2$
are the vacuum expectation values of two Higgs doublets.    
Apparently supersymmetry is broken and supersymmetric particles 
are of larger mass than their ordinary partners in the multiplet.
The mechanism of 'symmetry breaking' is unknown. Several
mechanism are proposed and introduce many new parameters~\cite{susybreak}.

\begin{figure}[htb]
    \hspace{-0.4cm}
    \includegraphics[width=0.58\textwidth]{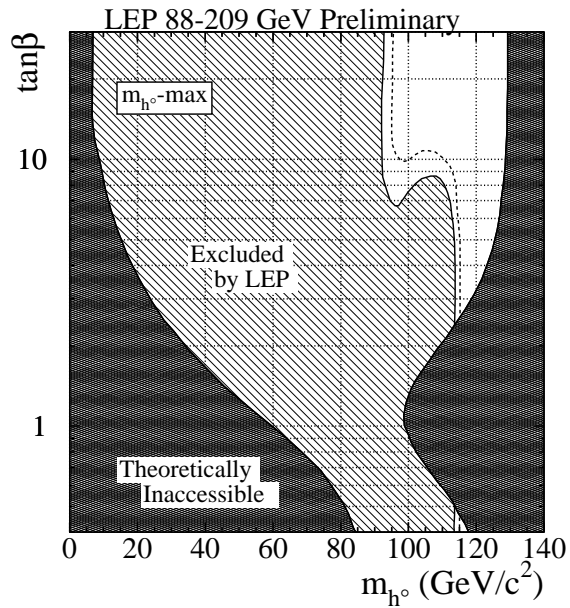}
    \caption[]{\label{fig:mssm}
             The excluded region in the $\Mh$-$\tan \beta$ plane
             in the '$\Mh$ max'. 
             The results of all LEP experiments are combined.
            } 
\end{figure}

In supersymmetric models the Higgs sector contains at
least
two scalar dublets~\cite{mssm}, resulting in 5 physical 
Higgs bosons.
Two neutral ones, h and H, are CP even, 
one neutral A is CP odd
and two, H$^\pm$, are charged.
Mixing between the two CP even
eigenstate introduces a mixing angle 
$\alpha$.
The production of the lightest Higgs boson h proceeds
either like in the SM in association with a Z or
in association with the CP odd A.
The couplings depend on $\alpha$ and $\beta$;
\begin{eqnarray*}
\frac{\hZZ}{\HZSM} & = & \sin(\beta - \alpha)\\
\frac{\hAZ}{\HZSM} & = & \cos(\beta - \alpha), 
\end{eqnarray*}
where $\HZSM = \frac{1}{4}(g^2+g'^2)\upsilon_0$ denotes 
the ZZH coupling in the SM.
The decay into the $\BB$ final state also depends on these parameters: 
\begin{eqnarray*}
\frac{\hBB}{\HbSM} & = & -\frac{\sin \alpha}{\cos \beta} \\
\frac{\ABB}{\HbSM} & = & \tan \beta ,
\end{eqnarray*}
where $\HbSM = \HBB$ is the Yukawa coupling.
The masses of the Higgs bosons are related to each other
and to the masses of the gauge bosons.
In particular, the mass of the lightest Higgs boson, $\Mh$,
is predicted to be smaller than $\MZ\cdot|\cos 2 \beta|$.
Radiative corrections, depending strongly
on the top-quark mass and the mixing 
of the scalar partners of the
left- and right-handed top, 
shift $\Mh$ to larger values~\cite{hollik}.
Both processes, $\EEhZ$ and $\EEhA$, are searched for at LEP 
using similar techniques as for the search of the SM Higgs boson.
No signal is found, and exclusion limits are
determined as function of $\tan \beta$. This is shown in
 Figure~\ref{fig:mssm} ~\cite{mssmlep} in the so called '$\Mh$ max' scenario,
in which the upper theoretical bound on
$\Mh$ becomes maximal.
Independent of $\tan \beta$, the mass limits for neutral Higgs
bosons in the MSSM are set to $\Mh > 91.0 \GeV$ and $\MA > 91.9 \GeV$.
No signal of charged Higgs bosons, charginos, neutralinos
or sfermions were found at LEP and limits on their masses
or production cross sections are set~\cite{lepsusy}. 

\begin{figure}
    \includegraphics[width=0.50\textwidth]{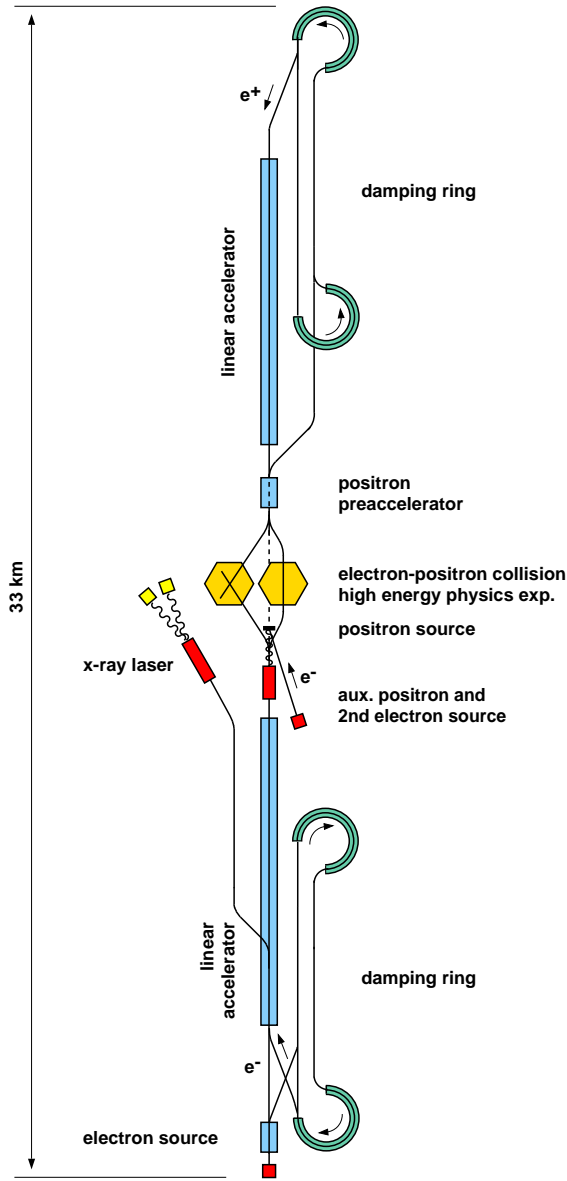}
    \caption[]{\label{fig:teslaacc}
            The layout of the TESLA linear collider at DESY   
            } 
\end{figure}%
\newpage

\stepcounter{section}
\section*{\large \arabic{section}.\ \large The TESLA project}  
 
The LEP collider approached the technical frontier
of a circular electron accelerator. 
The energy loss
of the accelerated electrons
per turn due to synchrotron radiation
is $\Delta {E} \simeq {E_b}^4/R$, 
where $E_b$ is the beam energy  and $R$ the radius of the 
accelerator. At the highest LEP 
energies these losses reached with about 3 \GeV~
per turn the limit of the RF power.
For a linear collider such losses do not exist. The main problems
here are the high gradient of the acceleration cavities, in order
to get the required energy with a technically reasonable length
of the accelerator, and the luminosity. Electron or positron bunches 
are brought into collision only once, hence the permanent creation
of new beam particles allowing high beam intensities is an issue.   
The advantages of an $\EE$ collider, in comparison to a proton machine,
are the well defined initial state and 
the possibility
to tune both $\rts$~ and the polarisation 
of electrons and positrons very precisely. Furthermore, 
also e$^-$ e$^-$, e$^- \gamma$ and $\gamma\gamma$ scattering 
are options to extend the physics potential.
A linear $\EE$
collider operated in the energy range below 1 TeV
would allow to study 
triple gauge boson couplings on a precision 
level sensitive to new physics,
would cover the threshold 
of $\EE \ra \TOP$ production, and in particular, would be the ideal
machine to explore the Higgs mechanism
or any new particle discovered in the new energy domain.
A sketch of the TESLA~\cite{tesla} linear collider, 
proposed by the DESY laboratory,
is shown in Figure~\ref{fig:teslaacc}.
The accelerator is foreseen to be constructed
starting from the DESY site in Hamburg
in north-west direction.
The total length will be about 30 km.
\begin{figure*}[t]
  \begin{center}
    \includegraphics[width=0.75\textwidth]{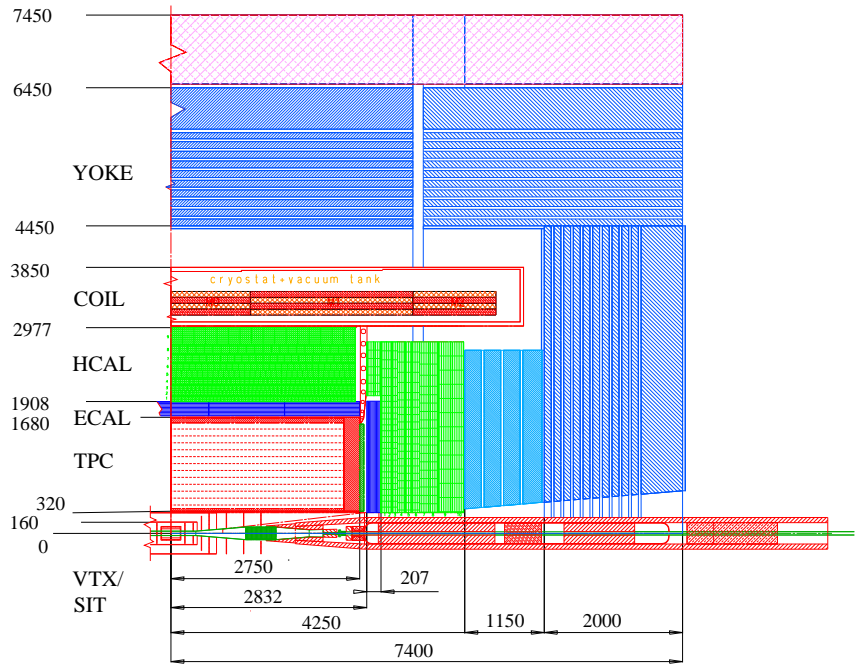}
    \caption[]{\label{fig:detector}
            The layout of the detector for TESLA.   
            } 
  \end{center}
\end{figure*}%
The basic parameters of TESLA are:
\begin{center}
\begin{table}[htb]
\begin{tabular}{lc}
$\sqrt s$           & $\le$ 0.8 TeV  \\
gradient            & 23.4 $\MeV$/m      \\
repetition rate     & 5  Hz         \\ 
beam pulse length   & 950  $\mu$s       \\ 
No. of bunches      & 2820       \\
per pulse           &          \\  
bunch spacing       & 337 ns        \\
charge per bunch    & 2$\cdot 10^{10}$        \\ 
beam size, $\sigma_x$& 553 nm       \\       
beam size, $\sigma_y$& 5  nm        \\      
bunch length          & 0.3 mm       \\  
Luminosity          & 3.4 10$^{34}$cm$^{-2}$s$^{-1}$ \\  
\end{tabular}
\end{table}
\end{center}
In the interaction region one detector for
the measurement of $\EE$ annihilations is foreseen. For a second
detector, designed to measure $\gamma\gamma$ collision,
an option exists. 
The feasibility of photon beams is under study.
The detector, of which a quarter
is depicted in Figure~\ref{fig:detector},
follows in general the structure of a LEP detector.
Starting from the interaction point there is a 
silicon tracker (VTX/SIT),
a Time Projection Chamber (TPC) as 
the main tracking device, an electromagnetic (ECAL) and a 
hadron (HCAL) calorimeter.
All these subdetectors are housed in a 
superconducting solenoidal
magnet of 3 T.
Much better in comparison to the LEP detectors
are, however, the performances of the subdetectors. 
The envisaged
momentum resolution
is
$\frac{\sigma_{p_t}}{p_t} = 4 \cdot 10^{-5} \cdot p_t~ [\GeV] $,
the impact parameter resolution 
$\sigma = 2.9 \oplus \frac{3.9}{p \cdot sin^{3/2} \theta)} \mu$m,
the resolution of the energy measurement in the electromagnetic
calorimeter  
$\frac{\sigma_{E_{el}}}{E_{el}} = \frac{11\%}{\sqrt E_{el}} + 0.6\% $
and in the hadron calorimeter
$\frac{\sigma_{E_{h}}}{E_{h}} = \frac{35\%}{\sqrt E_{h}} + 3 \%$ .

\stepcounter{section}
\section*{\large \arabic{section}.\ \large Physics} 

\begin{figure}[htb]
    \HS{-0.4}
    \includegraphics[width=0.53\textwidth]{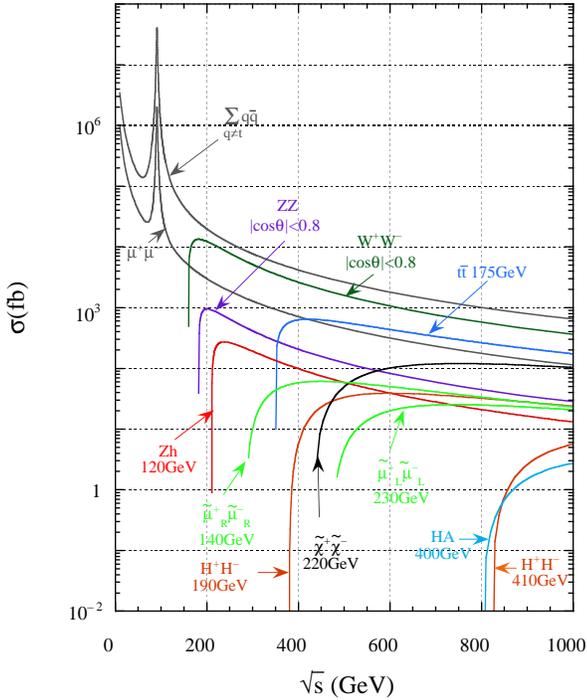}
    \caption[]{\label{fig:sigtot}
            The cross sections of  $\EE$ annihilations
            in several final states as expected in the 
            $\rts$ range of TESLA. Also shown are the expectations
            for the production of supersymmetric particles
            for given masses.
            } 
\end{figure}

\begin{figure}[htb]
\VS{-1}
  \begin{center}
    \includegraphics[width=0.5\textwidth,height=0.55\textwidth]{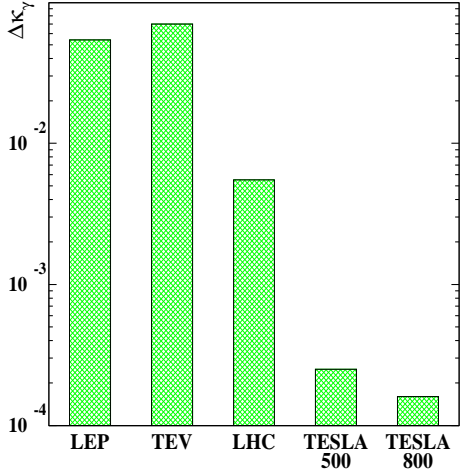}
    \caption[]{\label{fig:gaugecoup}
            The error on $\Delta\kappa_{\gamma}$ obtained at LEP  
            and the expectation for the LHC and TESLA.
            At TESLA estimates for $\rts = 500 \GeV$ and $\rts = 800                       \GeV$ are given
            ~\cite{tesla}.} 
  \end{center}
\end{figure}

The cross sections of $\EE$ annihilations 
in different final states as
function of the centre-of-mass energy 
are shown in 
Figure~\ref{fig:sigtot}.
Choosing \eg~ $\rts = 350 \GeV$
and an integrated Luminosity
of 500 fb$^{-1}$ in one year of running
the following event numbers are expected:
\begin{tabular}{lc}
process            &  No. of events   \\
$\EEEE$            &  1.5$\cdot$10$^8$  \\
$\EEWW$            &  5 $\cdot$10$^6$   \\
$\EEQQ$            &  3 $\cdot$10$^6$   \\
$\EEMM$            &  5 $\cdot$10$^5$   \\
$\EEZZ$            &  4 $\cdot$10$^5$   \\
$\EEZH$            &  8 $\cdot$10$^4$   \\ 
($\MH$ = 120 $\GeV$) &   \\
\end{tabular}  
With such a statistics many measurements
done at LEP can be performed 
at higher energy with considerably 
better accuracy~\cite{tesla}. Cross section measurements
of $\EEFF$ will be sensitive to
contact interactions up to several 10 TeV.
Up to the same mass scale extra dimensions~\cite{extra} can be probed.
The study of $\EEWW$ will allow to determine the triple 
gauge boson couplings, $\Delta\kappa$ and $\Delta g_1^Z$,
with an accuracy of $\sim 10^{-4}$. This is of particular interest, 
since radiative corrections to these quantities
in the framework of supersymmetry amount to $\sim 10^{-3}$.
A comparison of the accuracies obtained
for $\Delta\kappa_{\gamma}$ at LEP and expected 
at LHC and TESLA at $\rts = 500 \GeV$ and $\rts = 800 \GeV$
is shown in    
Figure~\ref{fig:gaugecoup}.
Another important issue is the study of
the process
$\EETOP$, which is accessible for the first time.
From a threshold scan, the top-quark mass, $\MTOP$,
will be determined with an accuracy of $\simeq 30 \MeV$.
Other topics are the couplings of the top-quark
to gauge bosons, the Lorentz-structure in top-quark
decays and the top-quark Yukawa coupling to the Higgs boson.
The principal subject will be, however, the Higgs boson.
The precision measurements at LEP and SLC 
on the Z point to a light Higgs boson, which will be accessible 
at TESLA. Due to the high event
statistics expected, detailed studies of the profile of
the Higgs boson will be possible. 
This is discussed now in more detail.

\stepcounter{section} 
\section*{\large \arabic{section}.\ \large Higgs Boson}

\begin{figure}[htb]
  \begin{center}
    \includegraphics[width=0.5\textwidth,height=0.45\textwidth]{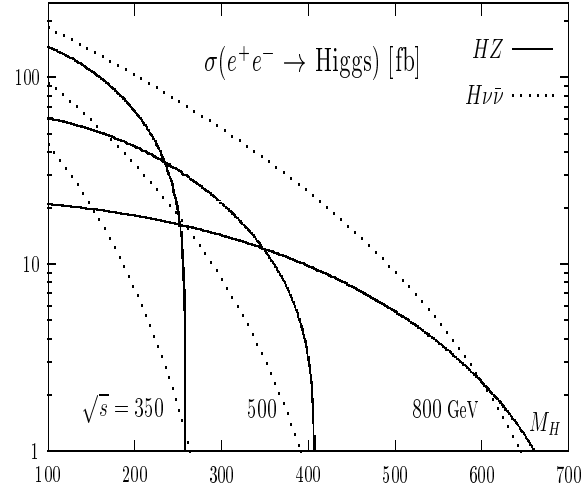}
    \caption[]{\label{fig:hprocs}
            The cross section of Higgs-strahlung and
            WW fusion as function of $\MH$ for
            $\rts = 350, 500$ and $800 \GeV$  
             ~\cite{tesla}.} 
  \end{center}
\end{figure}

\begin{figure*}[t]
  \begin{center}
    \includegraphics[width=\textwidth]{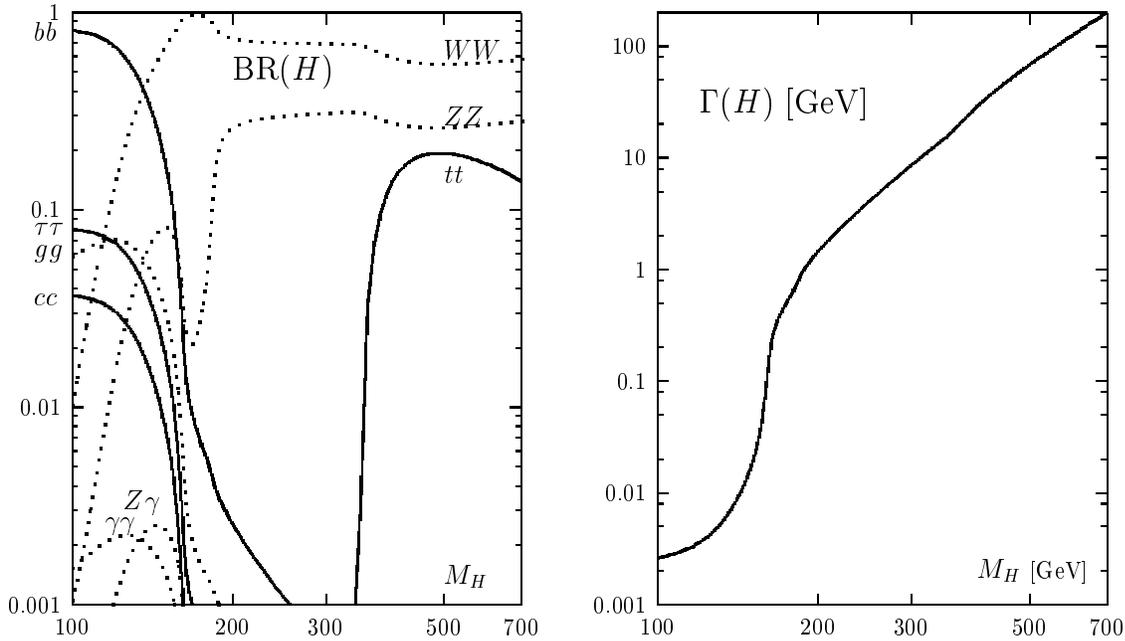}
    \caption[]{\label{fig:hibra}
            The branching fractions of the Higgs boson
            into fermions and bosons
            as function of $\MH$ (left). The total width
            of the Higgs boson as function of $\MH$ (right)
             ~\cite{tesla}.} 
  \end{center}
\end{figure*}

The cross section for the two dominant
processes of Higgs boson production,
the Higgs-strahlung and the WW fusion~\cite{higgsst}, read:
\begin{eqnarray*}
\sigma({\HZ}) = \frac{g_{\rm ZZH}^2\gf(v_e^2+a_e^2)}{96\sqrt{2} \pi  s}\\
             \times \beta \frac{\beta^2+12 \MZ^2/s}{(1-\MZ^2/s)^2}
\end{eqnarray*}
\begin{eqnarray*}
\sigma({\NUNUH}) \simeq \frac{g^2_{\rm WWH}}{4 \pi} 
                    \frac{\gf^2}{8 \pi^2}
                     ((1+ \frac{\MH^2}{s}) \\
                    \times\log \frac{s}{\MH^2} -
                    2(1-\frac{\MH^2}{s})),
\end{eqnarray*}
where $g_{\rm ZZH}= \frac{1}{4}(g^2+g'^2)\upsilon_0$ and $g_{\rm WWH} =
\frac{1}{2}g'^2\upsilon_0$.
These cross sections are
depicted as function of $\MH$ for several $\rts$
in 
Figure~\ref{fig:hprocs}.
The decay into fermions and bosons is described as:
\begin{eqnarray*}
\Gamma({\HTOF}) = \frac{g^2_{\HF}}{4 \pi} \frac{N_c}{2} \MH \\
                \times(1-4 \rm m_f^2/\MH^2)^{\frac{3}{2}}
\end{eqnarray*}
and
\begin{eqnarray*}
\Gamma({\HTOV}) = \frac{g^2_{\rm VVH}}{4 \pi} \frac{3}{8 \MH }
                (1-\frac{\MH^2}{3\rm m_V^2} \\
+\frac{\MH^4}{12\rm m_V^4})
                (1-4 \rm m_V^2/\MH^2)^{\frac{1}{2}}, 
\end{eqnarray*}    
where V = W or Z.
Using these formulae the branching fractions of the Higgs boson in
the several final states are calculated \footnote{The decays
$\HTOG$ and $\HTOGL$ are possible via quark loops.}
and shown in 
Figure~\ref{fig:hibra}.
For $\rts \simeq 350 \GeV$  and a $\MH$ range from 100 to 200~\GeV~  
the dominant production mechanism will be Higgs-strahlung
with substantial contribution from WW fusion, and
the dominant decay modes will be $\HTOB$ for $\MH \simeq $100 \GeV~ 
and $\HTOW$ for  $\MH \simeq $200 \GeV.
\stepcounter{subsection} 
\subsection*{\normalsize \thesubsection.\ Mass and Width}
\begin{figure}[htb]
  \begin{center}
    \includegraphics[width=0.5\textwidth]{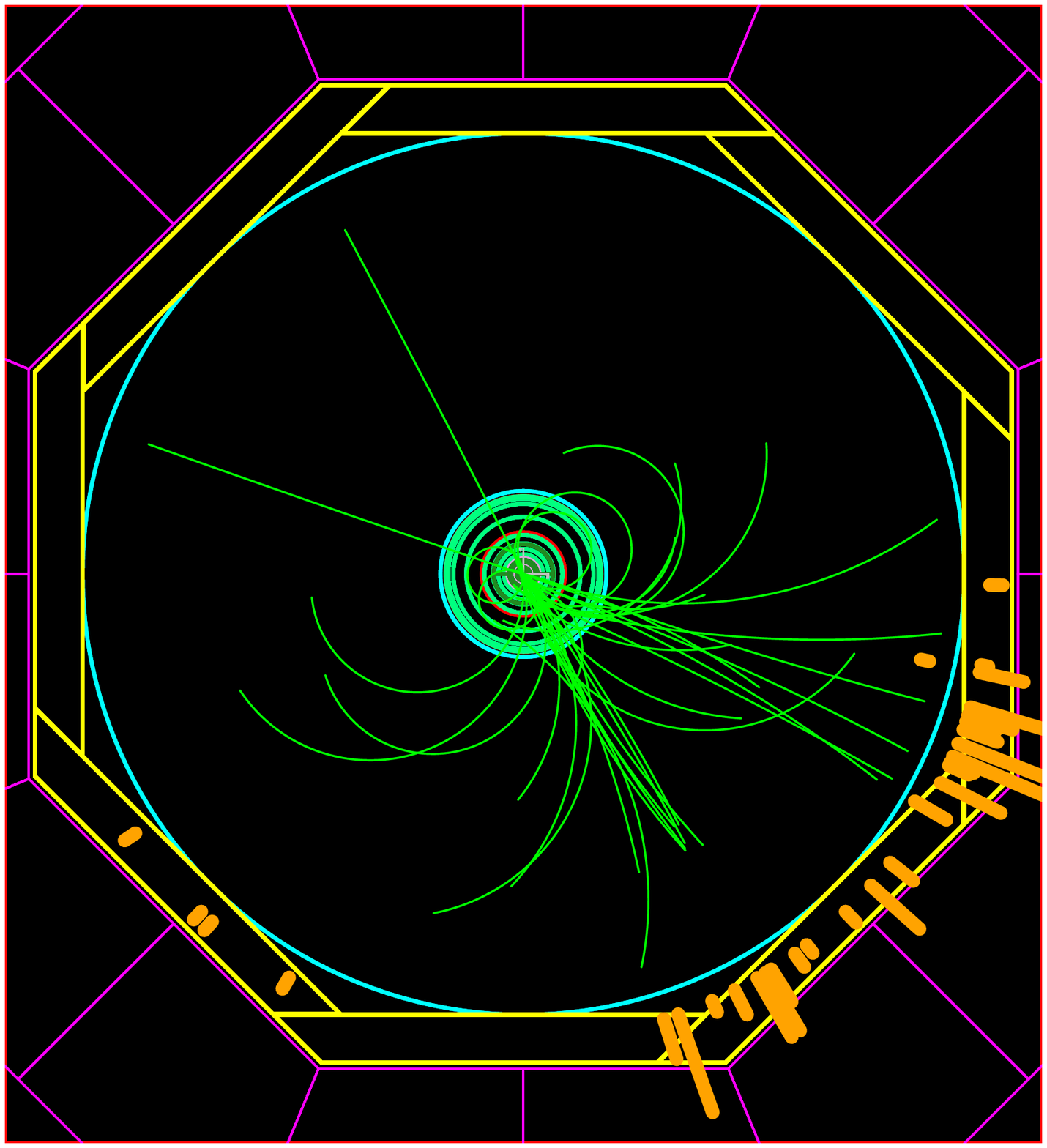}
    \caption[]{\label{fig:zreco}
            An event display for $\EEZH$ with $\ZMM$ and 
            $\HTOB$. The isolated tracks of the two muons
            and the two b-quark jets are clearly visible. 
            } 
  \end{center}
\end{figure}

\begin{figure}[htb]
  \begin{center}
    \includegraphics[width=0.5\textwidth]{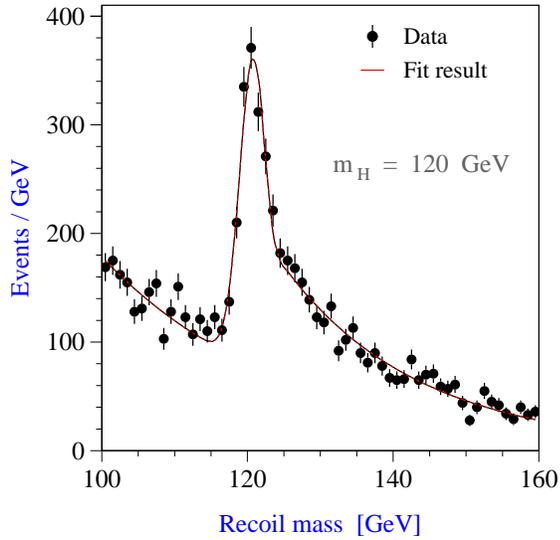}
    \caption[]{\label{fig:zrecospe}
             The distribution of the recoil mass
             against the Z in $\EEZH$ events
             and Z decays into electrons and muons
             assuming $\MH = 120 \GeV$.  
              } 
  \end{center}
\end{figure}

\begin{figure}[htb]
  \begin{center}
    \includegraphics[width=0.5\textwidth]{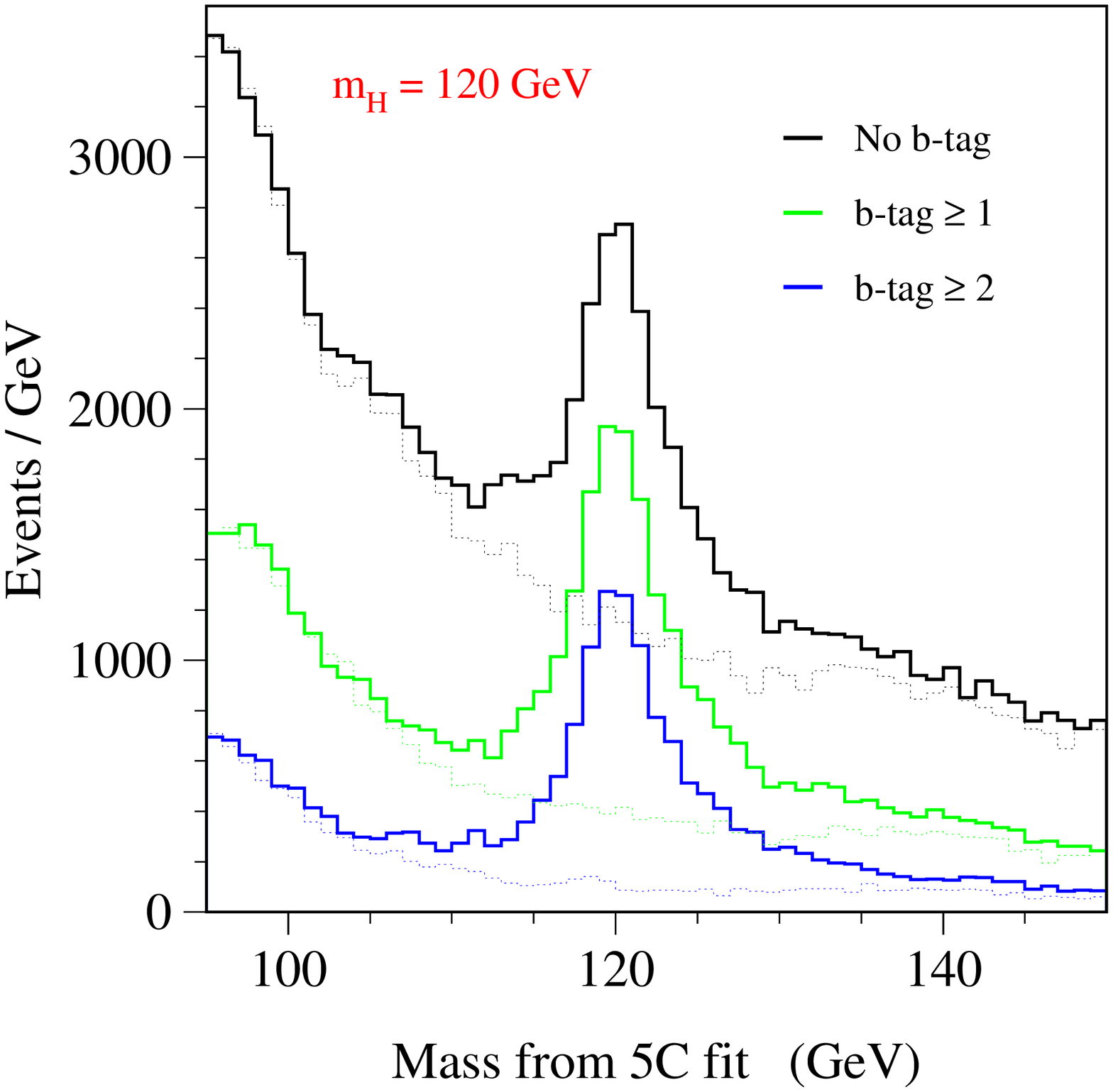}
    \caption[]{\label{fig:fourjet}
            The invariant mass distribution of the two
            b-quark jets from 
            $\HTOB$ after imposing
            energy and momentum conservation
            and constraining the mass of the two jets 
            assigned to the Z to $\MZ$
             ~\cite{tesla}.} 
  \end{center}
\end{figure}

\begin{figure}[htb]
  \begin{center}
    \includegraphics[width=0.5\textwidth]{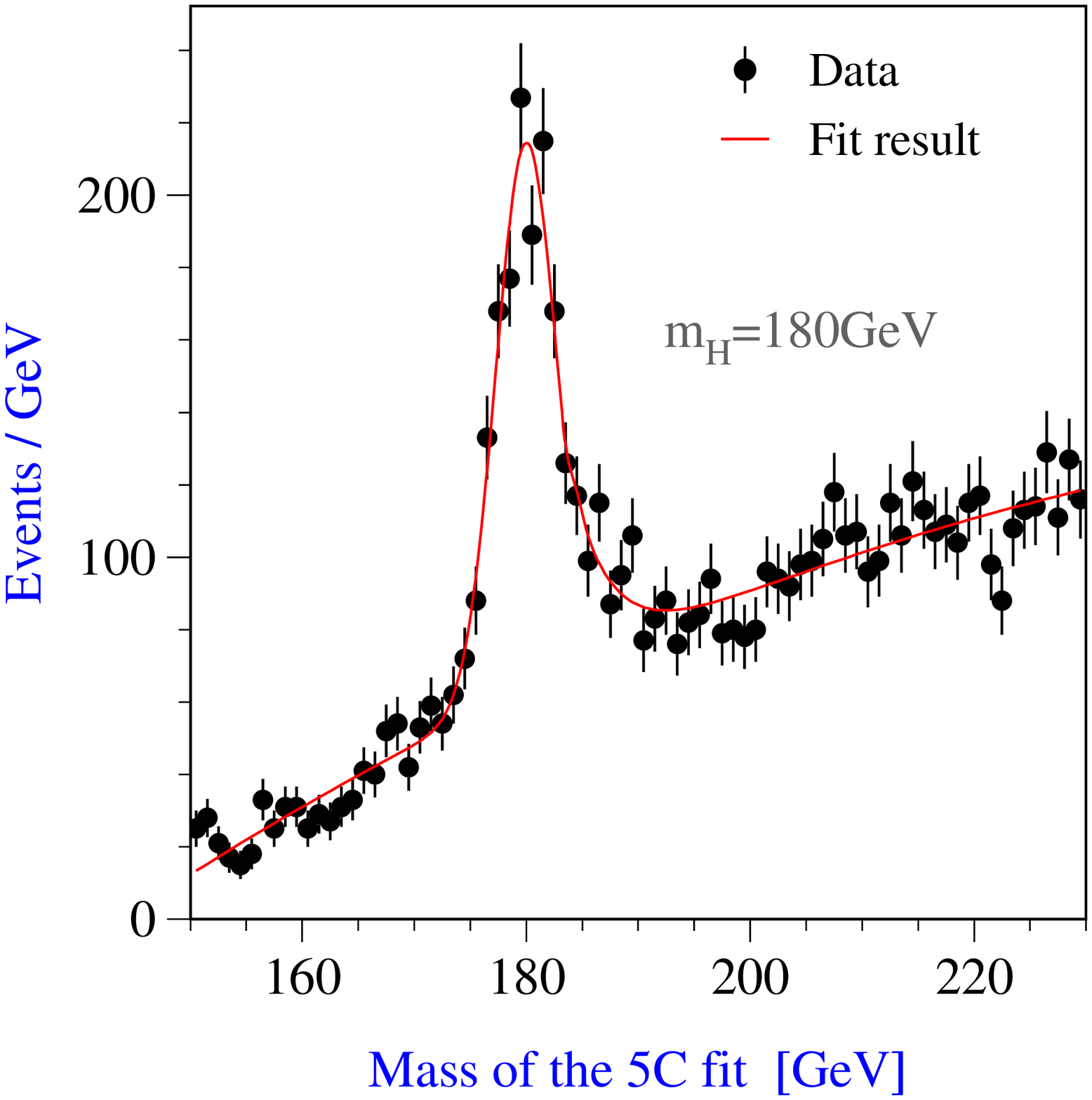}
    \caption[]{\label{fig:sixjet}
            The invariant mass distribution of the four jets
            stemming from  
            $\HTOW$ decays after imposing
            energy and momentum conservation
            and constraining the mass of the two jets 
            assigned to the Z to $\MZ$
             ~\cite{tesla}.} 
  \end{center}
\end{figure}

The first quantity to be determined
experimentally is clearly the Higgs mass.
The most model independent way to measure
$\MH$ is using the recoil mass against the Z.
The Z is identified by its decay into electrons and muons,
and the recoil mass is obtained as
\begin{eqnarray*}
{\rm m_{\mathrm{R}}}^2  =  s - 2 \cdot \sqrt s \cdot E_{\mathrm{Z}} + \MZ^2.
\end{eqnarray*}
$\rm E_{\mathrm{Z}}$ is the  energy of the Z 
reconstructed from the two leptons. 
Events of this topology have a very clear signature 
in the detector as shown in 
Figure~\ref{fig:zreco}.
The spectrum of the recoil mass, $\rm m_{\mathrm{R}}$, is displayed
in 
Figure~\ref{fig:zrecospe} assuming $\MH = 120 \GeV$.
The Higgs boson appears as a very sharp and 
clear signal 
over a small background. The latter originates mainly
from $\EEZZ$. The precision of $\MH$ obtained from
a fit to this distribution is
110 \MeV.  
The accuracy of the mass measurement can be improved 
by using also the hadronic decay channels
of the Z and performing a kinematic fit to the whole event
requiring energy and momentum conservation.
The invariant mass
distribution of the two jets 
assigned to the Higgs boson decay is shown in 
Figure~\ref{fig:fourjet}.
Also shown are the distributions obtained after the requirements
that one or two jets contain a secondary vertex caused by a 
b-quark decay.
This requirement supresses strongly the background 
whereas the signal stays nearly unchanged.
The precision obtained for $\MH$ 
is
45 \MeV.  
If $\MH$ approaches the $\WW$ threshold  the decay
$\HTOW$ becomes dominant and causes
more complex final states. Using proper jet clustering 
algorithms the decay channels  
$\zhllww$ and $\zhqqww$ can be reconstructed.
This is shown in Figure~\ref{fig:sixjet}
for the channel $\zhqqww$ with six hadronic jets in the final state.
The Higgs boson signal is very clear on top of a 
moderate background. The accuracy of the mass measurement
is $\simeq 110 \MeV$.
Combining all channels the Higgs boson mass
will be measured with an accuracy of $\simeq 40 \MeV$
at $\MH \simeq 120 \GeV$ and $\simeq 70 \MeV$ 
at $\MH \simeq 180 \GeV$.

The width of the Higgs boson, 
$\Gamma({\rm H})$
is of several $\GeV$ for masses $\MH > 250 \GeV$ and 
the detector performance allows to measure it directly
from the invariant mass distribution.
At lower masses, however, the mass resolution is
much larger than $\Gamma({\rm H})$.
But using the relation 
$\Gamma({\rm H}) = \frac{ \Gamma(\HTOX)}{\cal{B}(\HTOX)}$,   
with X = WW, ZZ or $\gamma \gamma$, 
the width can be determined indirectly.
For example, the partial width 
of $\HTOG$ can be measured via Higgs boson production
in a $\gamma \gamma$ collider. The branching fraction
for $\HTOG$ is accessible in the $\EE$ collider
by using the $\gamma \gamma$ invariant mass spectrum.
The same exercise can be done for
WW fusion and $\HTOW$. Depending on $\MH$, the width of the
Higgs boson can be determined with an accuracy between
4 and 10\%. 

\stepcounter{subsection}
\subsection*{\normalsize \thesubsection.\ Branching Fractions}
\begin{figure*}[t]
  \begin{center}
    \includegraphics[width=0.7\textwidth]{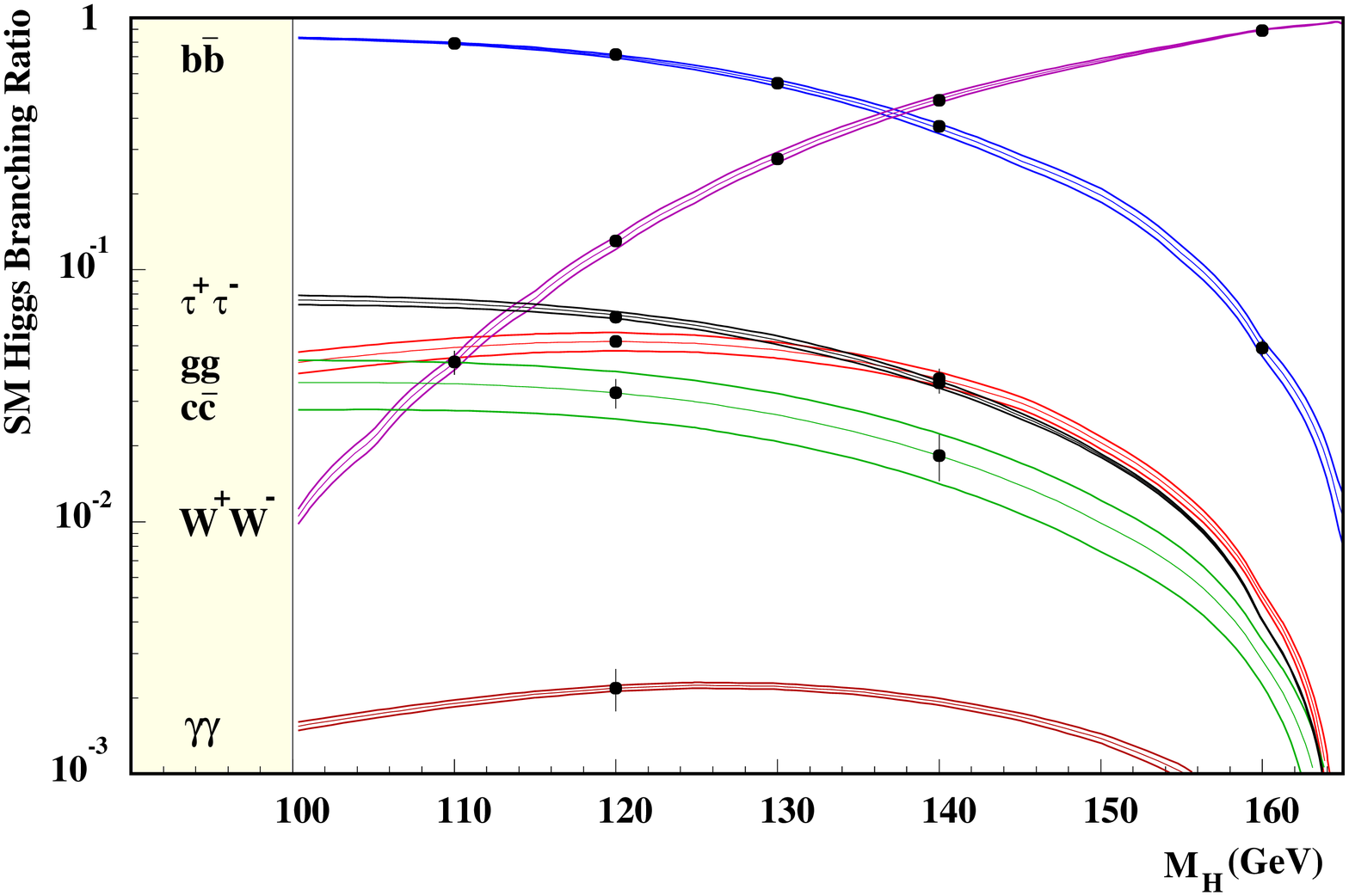}
    \caption[]{\label{fig:hibrameas}
            The measured branching fractions of the Higgs boson
            into fermions and bosons
            as function of $\MH$. Dots are measurements and
            the curves the predictions by the SM 
             ~\cite{tesla}.} 
  \end{center}
\end{figure*}

The branching fractions, as shown in  
Figure~\ref{fig:hibra},
are precisely predicted in the SM.
In supersymmetric extensions of the SM they 
are presumably different.
Hence its measurement
is crucial to
find out which structure of the Higgs sector
is realised in nature.
The measurement of branching fractions
makes use
of the mass distributions
shown in the previous section
and, in addition, uses the excellent flavour tagging
capabilities of the detector.
The result is shown in 
 Figure~\ref{fig:hibrameas}.
\stepcounter{subsection}
\subsection*{\normalsize \thesubsection.\ Spin}
\begin{figure}[htb]
  \begin{center}
    \includegraphics[width=0.5\textwidth]{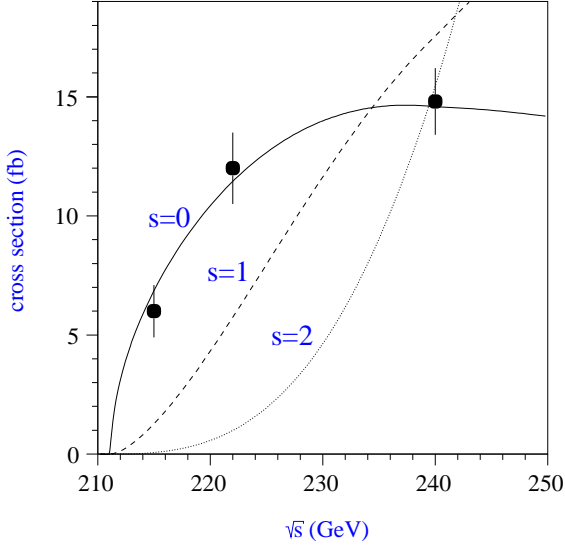}
    \caption[]{\label{fig:hspin}
            The cross section of $\EEZH \ra \elljet$ 
            just above the threshold assuming
            $\MH = 120 \GeV$. The dots correspond to
            a measurement and the curves are
            predictions
            for several spins 
             ~\cite{tesla}.} 
  \end{center}
\end{figure}

If a signal is seen in the dijet invariant mass distribution
the measurement of the spin is 
crucial for its identification as the Higgs boson.
It can be performed by analysing the 
energy dependence of the Higgs boson production 
cross section just above the kinematic
threshold. 
For a spin zero particle
the rise of the cross section is expected to be 
$\sim \beta$, where $\beta$ is the velocity of the boson
in the centre-of-mass system
\footnote{There are particular scenarios
for s=1 and 2, which show a threshold
behaviour similar in shape to the s=0 one. This can be disentangled 
using angular information in addition.}. 
For a  spin one particle the rise is $\sim \beta^3$
and for spin two like $\sim \beta^5$.
With a very small luminosity of about
20 fb$^{-1}$ per energy point the scalar nature of 
the Higgs boson can be established, as shown in
Figure~\ref{fig:hspin}.
\stepcounter{subsection}
\subsection*{\normalsize \thesubsection.\ Self Couplings}
\begin{figure}[htb]
  \begin{center}
    \includegraphics[width=0.5\textwidth,height=0.16\textwidth]{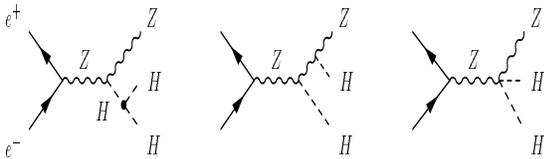}
    \caption[]{\label{fig:hself}
            The Feynman-diagrams 
            describing the triple Higgs self coupling (left) 
            and two background contributions of similar
            topology (right) 
            .} 
  \end{center} 
\end{figure}

\begin{figure}[htb]
  \begin{center}
    \includegraphics[width=0.5\textwidth]{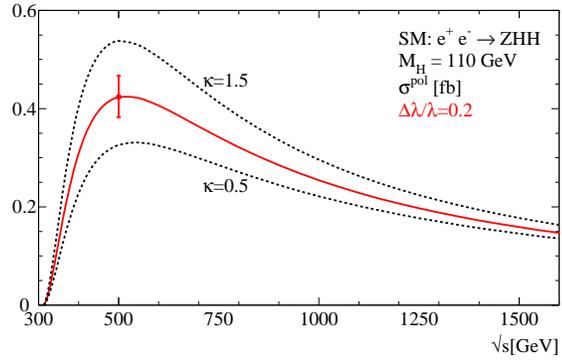}
    \caption[]{\label{fig:hselfm}
            The energy dependence
            of the cross section of $\EE \ra \rm{ZZH}$
            for $\MH=110\GeV$.
            The dot represents
            a measurement, the full line the expectation
            from the SM and the dashed line the
            prediction obtained if the SM cross section
            of the triple Higgs coupling contribution
            is multiplied by $\kappa$ 
             ~\cite{tesla}.} 
  \end{center} 
\end{figure}

From the potential in the Lagrangian
 terms remain which describe triple and 
quartic self couplings of the Higgs boson.
The measurement of these couplings would 
specify the Higgs potential containing 
the parameters $\upsilon_0$
and $\lambda$.
The triple Higgs boson couplings appear in the 
Feynman-diagram depicted in
Figure~\ref{fig:hself}.
At TESLA, asuming $\rts = 500 \GeV$ and 
an integrated luminosity of 1 ab$^{-1}$,
the triple Higgs boson couplings can be detected.
This is demonstrated in 
Figure~\ref{fig:hselfm}.
A measurement of $\lambda$ is feasible at
$\rts = 500 \GeV$ 
with an accuracy of $\simeq$ 20\%.

\stepcounter{section}
\section*{\large \arabic{section}.\  Summary}
From measurements at LEP the parameters of the 
heavy neutral gauge boson, Z, were precisely determined.
The SM was confirmed on the level 
of quantum corrections.
At high LEP energies the triple
gauge boson couplings,
predicted by the non-abelian structure of the
SM, were detected
and found to be in agreement with the prediction.
The Higgs boson, the missing key-stone of the SM, was not 
found. But from direct searches and from virtual Higgs 
boson contributions to observables 
its mass is estimeted to be between
114 and $\simeq$ 200 \GeV.
The TESLA collider offers an excellent physics program
to confirm or disprove the SM
in a new energy domain. In particular the
profile of the Higgs particle, its mass, spin,
and couplings, will allow to explore the 
mechanism of spontaneous symmetry breaking. 
The top-quark, the heaviest
elementary particle known so far, will be subject
to detailed studies.
Precision measurements on fermion couplings to gauge bosons,
triple and also quartic gauge boson 
self-couplings
will give us hints for new physics beyond
the SM. And of course, a new energy domain will be explored,
eventually being full of unexpected phenomena.

\section*{\large Acknowledgements}
I would like to thank the organisers of the school
for their kind hospitality and K. M\"onig, A. Raspereza,
S. Riemann and A. Stahl for the critical reading of the
text.


\end{document}